# Mapping gradient-driven morphological phase transition at the conductive domain walls of strained multiferroic films

M. J. Han[1,2]†, E. A. Eliseev[3]†, A. N. Morozovska[4]*, Y. L. Zhu[1], Y. L. Tang[1], Y. J. Wang[1], X. W. Guo[1,5], X. L. Ma[1,6]*

[1]Shenyang National Laboratory for Materials Science, Institute of Metal Research, Chinese Academy of Sciences, Wenhua Road 72, 110016 Shenyang, China

[2]University of Chinese Academy of Sciences, Yuquan Road 19, 100049 Beijing, China

[3]Institute for Problems of Materials Science, National Academy of Sciences of Ukraine, Krjijanovskogo 3, 03142 Kyiv, Ukraine

[4]Institute of Physics, National Academy of Sciences of Ukraine, 46, pr. Nauky, 03028 Kyiv, Ukraine

[5]University of Science and Technology of China, Jinzhai Road 96, 230026 Hefei, China

[6]State Key Lab of Advanced Processing and Recycling on Non-ferrous Metals, Lanzhou University of Technology, Langongping Road 287, 730050 Lanzhou, China

†These authors contributed equally to this work.

*Correspondence and requests for materials should be addressed to X. L. Ma (email: xlma@imr.ac.cn) or A. N. Morozovska (email: anna.n.morozovska@gmail.com)

## Abstract

The coupling between antiferrodistortion (AFD) and ferroelectric (FE) polarization, universal for all tilted perovskite multiferroics, is known to strongly correlate with domain wall functionalities in the materials. The intrinsic mechanisms of domain wall phenomena, especially AFD-FE coupling-induced phenomena at the domain walls, have continued to intrigue the scientific and technological communities because of the need to develop the future nano-scale electronic devices. Over the past years, theoretical studies often show controversial results, owing to the fact that they are neither sufficiently nor directly corroborated with experimental evidences. In this work, the AFD-FE coupling at uncharged 180° and 71° domain walls in $BiFeO_3$ films are investigated by means of aberration-corrected scanning transmission electron microscopy with high resolution (HR-STEM) and rationalized by phenomenological Landau-Ginsburg-Devonshire (LGD) theory. We reveal a peculiar morphology at the AFD-FE walls, including kinks, meandering, triangle-like regions with opposite oxygen displacements and curvature near the interface. The LGD theory confirms that the tilt gradient energy induces these unusual morphology and the features would change delicately with different kinds of domain walls. Moreover, the 180° AFD-FE walls are proved to be conductive with an unexpected reduction of Fe-O-Fe bond angle, which is distinct from theoretical predictions. By exploring AFD-FE coupling at domain walls and its induced functionalities, we provide exciting evidences into the links between structural distortions and its electronic properties, which benefit a lot for fundamental understanding for domain wall functionalities as well as functional manipulations for novel nano-devices.

## Introduction

Multiferroics, with simultaneous coexistence of ferroic long-range orders and accompanied unusual physical properties, are regarded as the fertile systems for condensed matter physics and multifunctionalities. As a result, exploring the interplay among these order parameters of different nature, such as ferroelectric (**FE**) polarization, antiferrodistortive (**AFD**), ferromagnetic (**FM**) or antiferromagnetic (**AFM**) orders has been regarded as a key issue both for fundamental studies and future applications[1-3]. Meanwhile, domain walls, due to their recently discovered manifold unusual physical properties, offer us an excellent platform for the corresponding investigations[4-6].

Domain walls are reported to perform series novel functionalities, including large local conductivity[4], photovoltaic effect[7,8], magnetoelectric response[9,10] and even superconductivity[11], which confirm that the wall itself can act as an active element in functional devices. Except for the exciting functionalities, the hidden physical mechanisms underlying in domain walls responsibility for the interacting and competing lattice, charge, spin and orbital degrees of freedom still require fundamental investigations[12,13]. Surprisingly, all these physical mechanisms and properties can be traced to the changes of the fundamental unit in transition metal oxides with an $ABO_3$ perovskite structure — the $BO_6$ octahedra. The arrangements or symmetry of the corner-sharing octahedral network in perovskites are reported to be strongly coupled with the electronic, magnetic and optical properties[14,15]. Investigations on the AFD characteristics and the coupling of AFD, FE and FM order parameters at domain walls would help us better understand the structure of the walls, permit in the explanation of wall multiferroicity in terms of their underlying structure and make it more accessible to control the specific functionalities. For instance, Catalan et al.[16,17] reported that the local conduction may be generated by the considerably reduced electronic bandgap that corresponds to the straightening of the octahedral rotation at the domain walls using first-principles calculations. Besides, first-principles calculations also shows that the changes in the Fe-O-Fe bond angles in $BiFeO_3$ at the domain walls may change the canting of the Fe magnetic moments accordingly, and end up with enhanced local magnetization at the domain walls[18]. Moreover, the interplay between strain, FE, AFE and AFD long-range order parameters would generate a lot of remarkable functionalities, including flexoelectric, rotostriction and flexoantiferrodisortive effects through calculations[19-22].

Compare to the abundant theoretical results reported on FE and AFD coupling at domain walls, experimental studies of the coupling are quite rare. Previous experimental studies on domain walls are mostly focused on the FE nature of domain walls. The AFD characteristics, especially the AFD-FE coupling at domain walls are rarely addressed. One reason is that it is still a challenge to probe both the AFD and FE characteristics locally at the domain walls, though the local probing is urgent for understanding of the AFD-FE interplay. Possible causes may be the large lattice and octahedra distortions at the domain walls which make it harder to probe the FE and AFD characteristics at the atomic scale. Except this, the out-of-phase rotation of oxygen octahedra in $BiFeO_3$, the only multiferroics reported at room temperature[23], makes it rather difficult to detect the rotation angle along $[100]_{pc}$ direction. Borisevich et al.[24] reported an atomically abrupt octahedral tilt transition across charged domain walls compared with diffuse associated polarization profile through principle component analysis based on simultaneously acquired HAADF and ABF images. However, the coupling between FE and AFD at domain walls and its influence on the local properties are far from being settled.

In this work, we used HR-STEM to acquire high-angle annular dark-field (**HAADF**) and annular bright-field (**ABF**) images at 180° and 71° domain walls in BFO films and extract the FE polarization and AFD characteristics locally around the domain walls directly from the ion positions in the images. An unusual AFD-FE wall morphology including kinks, meandering, triangle-like regions with opposite oxygen displacements and curvature near the interface are observed. Landau-Ginsburg-Devonshire (**LGD**) theory based calculations confirm that the meandering domain walls originates from the tilts gradient energy in thin BFO films, whereas the curvature near the interface at 71° domain walls is induced by the surface influence. Most importantly, we observed an obvious conductivity accompanied with a distinct decrease of Fe-O-Fe bond angle at 180° domain walls which is completely different from previous reports. Thus, for the first time, we provide direct experimental evidences for AFD-FE coupling and put forward that the correlated conduction may ascribed to the reduction of Fe-O-Fe bond angle at 180° domain walls, which may enlighten the study on corresponding electronic properties.

## Results

### Experimental results for AFD-FE walls

The AFD order parameter corresponds to the rotations of oxygen octahedra as shown in Fig.1a. The deformation or rotation of the oxygen octahedra can give rise to new functionalities. Particularly, ferroelectric domain walls would generate changes in polarization and oxygen octahedra at the same time. The coupling between them ends up with different AFD-FE walls in ferroelectric films (see Fig. 1b-c). To demonstrate the AFD and FE coupling at the domain walls, we study the 180° and 71° walls in $BiFeO_3$ (BFO) films with different thicknesses grown on $PrScO_3$ (PSO) $(001)_O$ substrates (see Fig. S1). Since the oxygen octahedra in BFO share an out-of-phase rotation along $[100]_{pc}$ direction, in order to demonstrate the AFD walls and FE walls at the same time, we managed to acquire the HAADF and ABF images for cross-sectional BFO films project along $[110]_{pc}$ direction.

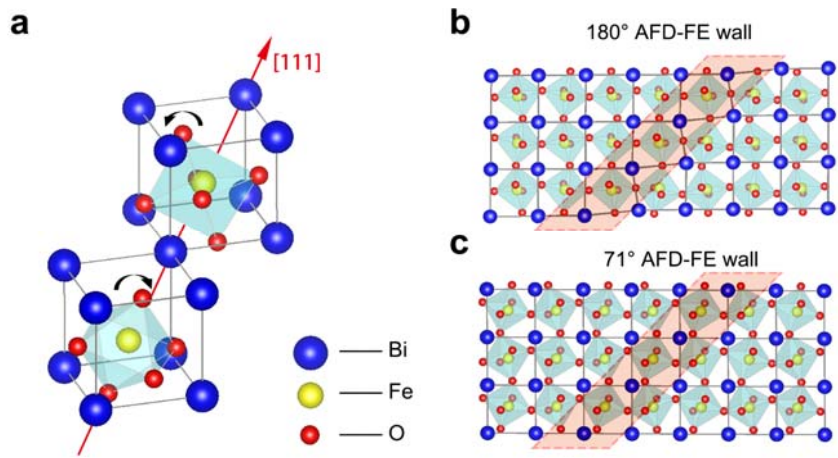

**Figure 1| Crystal structures showing the antiferrodistortion and AFD-FE walls in $BiFeO_3$ film.** (a) Atomic structure of bulk $BiFeO_3$ crystal showing the oxygen octahedral rotation (black curved arrows). (b-c) Atomic structure models for 180° (b) and 71° (c) AFD-FE walls.

The $(\bar{1}10)$ oriented 180° domain walls in a 18 nm BFO film grown on PSO substrate are shown in Fig. 2. The cross-sectional HAADF image of BFO film (see Fig. 2a) reveals two 180° domain walls in it. The enlarged images, which reveal clear opposite $Fe^{3+}$ ion displacements, are shown as the inset in each domain. The polarization directions (opposite to the $Fe^{3+}$ ion displacements) are also marked out by the yellows arrows in Fig. 2a. Detailed analyses on the 180° domain walls in the red and blue rectangles are shown in Fig. 2b-c, respectively. The 180° FE domain walls are quite straight except for several kinks (marked out by the red arrows). Compare to this, the 180° AFD walls are much more tortuous based on the oxygen octahedra displacements extracted from ABF images of the domain walls (see Fig. 2d). Despite the opposite polarization directions, all of the domains demonstrate a checkerboard pattern of oxygen displacements, which can be clearly seen in the insets in Fig. 2d. Fig. 2e and 2f are oxygen displacements along in-plane and out-of-plane directions, respectively, which reveal a significant discrepancy between the FE and AFD walls. Particularly, there exist two triangle-like regions with opposite oxygen displacements (the red and blue areas marked out by white dashed lines) at each domain wall in Fig. 2e. It seems that the different oxygen displacement regions are related to the polarization direction in the two domains. The slightly smaller oxygen displacements at the interface than the ones at the surface are supposed to come from the large tensile strain from the substrate. Besides this, the oxygen displacements along out-of-plane direction demonstrate a significant meandering of the AFD walls (marked out by the white dashed areas in Fig. 2f). There also exist small meandering areas at each domain wall, which are marked out by the white arrows in Fig. 2f.

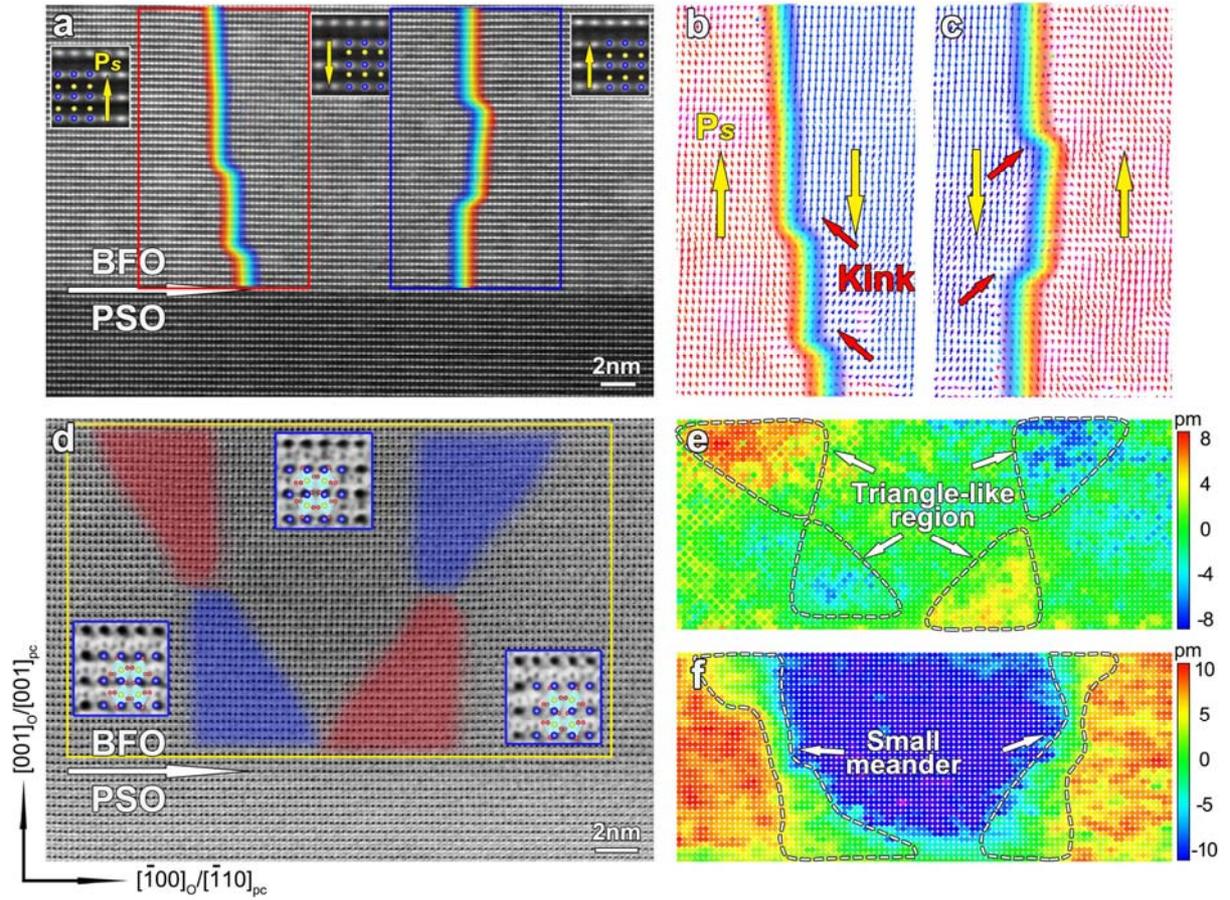

**Figure 2| 18 nm BiFeO₃ film grown on PrScO₃ (001)$_O$ substrate (180° domain wall)** (a) cross-sectional HAADF image of BFO/PSO interface project along [110]$_{pc}$ direction. Two 180° domain walls are marked out. The insets are enlarged images of each domain. The blue and yellow solid circles denote the position of $Bi^{3+}$ and $Fe^{3+}$ columns. The yellow arrows denote the polarization direction in each domain. (b) and (c) are $Fe^{3+}$ displacements mapping of the red and blue rectangle in (a), respectively. The kinked domain walls are marked out by red arrows. (d) ABF image of the same area as image (a). The red and blue triangle-like regions are marked out to demonstrate the AFD walls. The blue insets are superposition of the atomic schematics with enlarged images of each domain. Images (e) and (f) are oxygen displacement components along [$\bar{1}$10]$_{pc}$ and [001]$_{pc}$, respectively. The triangle-like regions and AFD wall meandering are marked out by white dotted areas.

We perform *in situ* piezoresponse force microscopy (PFM) and conductive atomic force microscopy (CAFM) measurements on the cross-sectional samples to probe electronic properties of the domain walls (see typical example in Fig. 3). The 180° domain walls (vertical PFM response is shown in Fig. 3a) reveal a significant conductivity (CAFM response is shown in Fig. 3b). It is reported that the high conductivity can be related with the strengthening of Fe-O-Fe bond angle, which would give rise to reduced electronic bandgap at domain walls[16,17]. However, the strengthening of Fe-O-Fe bond angle has not been proved from our experiments. In contrast to this, the Fe-O-Fe bond angles shown in Fig. 3c, which were deduced from the O and Fe ion column positions in Fig. 2d, reveal a distinct decrease in octahedral rotation angle that is opposite to previous theoretical studies. However, the B-O bond length does not show any obvious changes at the 180° domain walls (see Fig. 3d). We tend to believe that the reduction of B-O-B bond angle is caused by the AFD-FE coupling, instead of the B-O-B bond angle strengthening, that give rise to the conduction at domain walls.

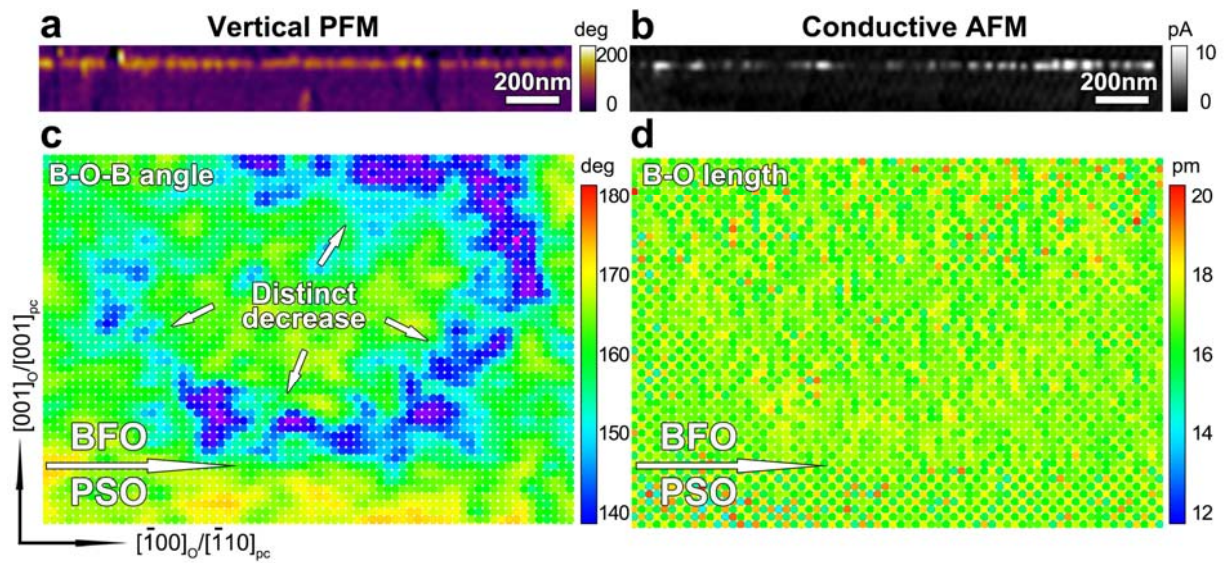

**Figure 3| Conductive AFM and corresponding B-O-B bond analyses for 180° domain walls.** (a-b) Vertical PFM (a) and corresponding conductive AFM (b) images for 180° domain walls. (c-d) B-O-B bond angle (c) and B-O bond length (d) distributions corresponding to the 180° domain walls in Fig. 2.

The 71° domain walls in 7 nm BFO film grown on PSO substrate are also analyzed for further investigation of AFD-FE coupling at domain walls (see Fig. 4). The polarization directions (opposite to the $Fe^{3+}$ ion displacements) marked out by the yellows arrows in Fig. 4a demonstrate a clear 71° domain wall. Detailed analysis on the 180° domain walls inside the red rectangle is shown in Fig. 4b. It can be seen that the 71° FE domain walls bend at the interface and the surface, as distinct from the 180° domain walls. The ABF image of the 71° domain wall (shown in Fig. 4c) demonstrates the coexistence of uniform spatial distribution (left) and a checkerboard pattern (right) of oxygen displacements. Fig. 4d is oxygen displacements along in-plane direction, which demonstrate the increased width of AFD wall compared to FE wall and significant curvature at the interface and surface (marked out by the white arrows in Fig. 4d).

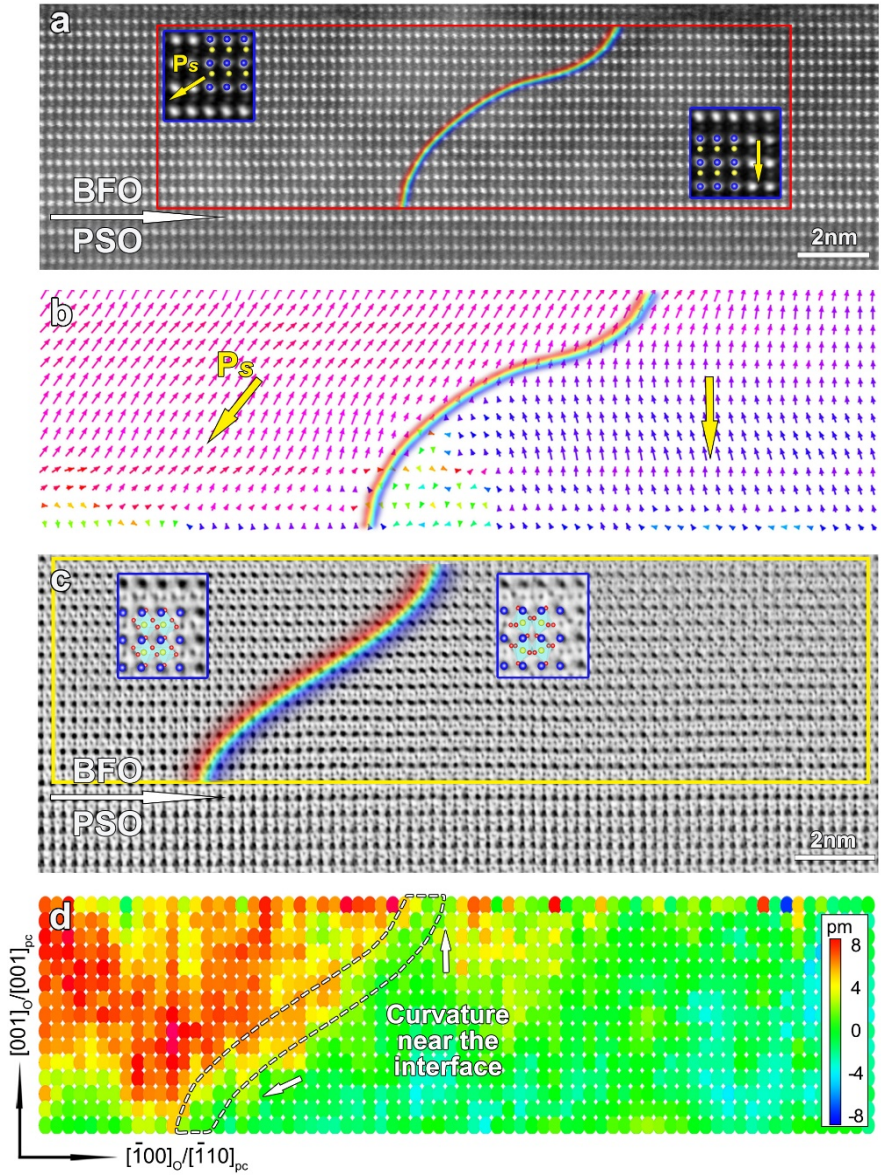

**Figure 4| 7 nm BiFeO₃ film grown on PrScO₃ (001)_O substrate (71° domain wall)** (a) HAADF image of BFO/PSO interface. The 71° domain wall is marked out. The insets are enlarged images of each domain. The blue and yellow solid circles denote the position of $Bi^{3+}$ and $Fe^{3+}$ columns. The blue arrows denote the polarization direction in each domain. (b) $Fe^{3+}$ displacements mapping of the red rectangle in figure (a). (c) ABF image of the 71° domain wall. The 71° AFD wall is marked out. The blue insets are superposition of the schematics with enlarged images of each domain. (d) Oxygen displacement components along $[\bar{1}10]_{pc}$ direction. The AFD wall broadening are marked out by white dotted areas. The white arrows denote the curvature near the interfaces.

**The model and parameters for theoretical description.**

Schematics of the considered system, consisting of a thin BFO film, namely its {111}-cut with incline domain stripes, placed between electrically conducting top and bottom electrodes, and PSO substrate are shown in Fig. 5. The considered nominally uncharged 180° and 71° domain walls (DWs) in BFO are shown in Fig. 1b-c. To simplify the consideration of 71° DWs we changed the coordinate frame from pseudocubic to orthorhombic settings by the rotation around $X_2$ axis on $\pi/4$ angle.. The two-dimensional (2D) problem was solved numerically by finite element modeling (**FEM**). Initial distributions of tilt and polarization components were taken in the form of 180° or 71° straight domain structures superimposed on the random distribution with a very small amplitude.

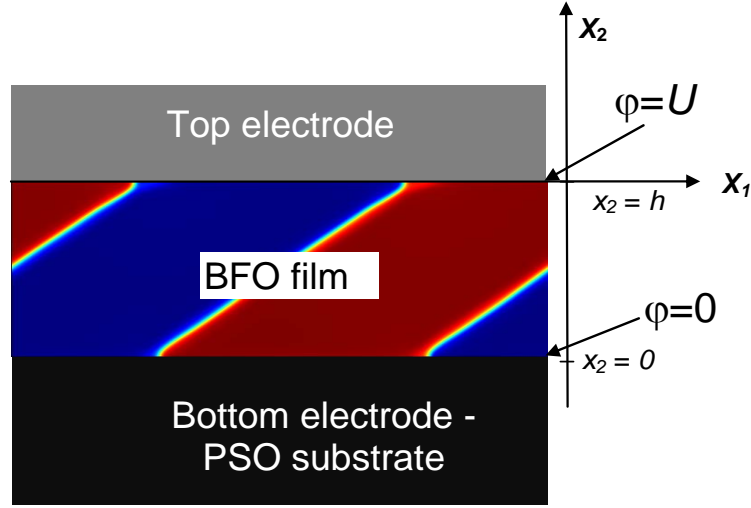

**Figure 5| Considered system for modelling.** Considered system, consisting of thin BFO film, placed between electrically conducting top and bottom electrodes, and PSO substrate. The {111}-cut with a domain stripes is shown.

We use a phenomenological LGD theory to model the domain walls (DW) morphology in a thin BFO film placed on a thick PSO substrate. Two vectorial order parameters, namely polarization components $P_i$ and oxygen octahedral tilts $\Phi_i$ are considered for the description of ferroelectric (FE) and antiferrodistortive (AFD) components of the DWs ($i$=1, 2, 3). The bulk part of LGD thermodynamic potential consists of the following contributions:

$$G_V = \int_S dx_1 dx_3 \int_0^h (\Delta G_{AFD} + \Delta G_{FE} + \Delta G_{BQC} + \Delta G_{striction} + \Delta G_{elast} + \Delta G_{flexo}) dx_2 \quad (1)$$

In Eq.(1) all contributions are consistent with the parent phase m3m symmetry in accordance with the basics of LGD approach, and its full form is listed in Appendix A in the Suppl. Mat. The compact form of the AFD contribution is:

$$\Delta G_{AFD} = b_i(T)\Phi_i^2 + b_{ij}\Phi_i^2\Phi_j^2 + b_{ijk}\Phi_i^2\Phi_j^2\Phi_k^2 + v_{ijkl}\frac{\partial \Phi_i}{\partial x_k}\frac{\partial \Phi_j}{\partial x_l} \quad (2a)$$

In accordance with the classical Landau approach, we assume that the coefficients $b_i$ are temperature dependent. In accordance with experiments[26], the dependence can be described by a Barrett law[25], $b_i = b_T T_{q\Phi}(\coth(T_{q\Phi}/T) - \coth(T_{q\Phi}/T_\Phi))$, where $T_\Phi$ is the AFD transition temperature and $T_{q\Phi}$ is a characteristic temperature. Description and numerical values of the phenomenological coefficients $b_i$, $b_{ij}$, $b_{ijk}$ and gradient coefficients $v_{ij}$ included in Eq.(2a) can be found in Table I, where Voight notations are used.

The compact form of the FE contribution is:

$$\Delta G_{FE} = a_i(T)P_i^2 + a_{ij}P_i^2 P_j^2 + a_{ijk}P_i^2 P_j^2 P_k^2 + g_{ijkl}\frac{\partial P_i}{\partial x_k}\frac{\partial P_j}{\partial x_l} - P_i E_i \quad (2b)$$

In accordance with LGD approach that is well-adopted for proper and incipient ferroelectrics, the coefficients $a_k$ are temperature dependent and obeys the Barrett law, $a_k^{(P)} = \alpha_T(T_{qP}\coth(T_{qP}/T) - T_C)$, where $T_C$ is the Curie temperature and $T_{qP}$ is a characteristic temperature[21, 25,26]. Description and numerical values of the phenomenological coefficients $a_i$, $a_{ij}$, $a_{ijk}$ and gradient coefficients $g_{ij}$ included in Eq.(2b) can be found in Table I. Electric field components $E_i$ are defined via electrostatic potential in the conventional way, $E_i = -\partial\varphi/\partial x_i$.

The potential satisfies Poisson equation, $\varepsilon_0 \varepsilon_{eff} \frac{\partial^2 \varphi}{\partial x_i^2} = \frac{\partial P_i}{\partial x_i}$, where the effective dielectric permittivity, $\varepsilon_{eff} = \sum_i \varepsilon_{bi} + \varepsilon_{el}$, includes a background permittivity[27], Jahn-Teller modes and electronic contributions, which in total can be pretty high for BFO. Electric boundary conditions for the short-circuited film are the zero electric potential at the conducting electrodes, $\varphi|_{x_2=0,h} = 0$.

The compact form of the biquadratic coupling energy between polarization and tilt is

$$\Delta G_{BQC} = \zeta_{ijkl} \Phi_i \Phi_j P_k P_l, \tag{2c}$$

As one can see, the coupling energy (2c) includes poorly known tensorial AFD-FE biquadratic coupling coefficients $\zeta_{44}$, $\zeta_{11}$ and $\zeta_{12}$ [33], which have been treated as fitting parameters to experiment (listed in Table I).

Electrostriction and rotostriction contributions are

$$\Delta G_{striction} = -Q_{ijkl} \sigma_{ij} P_k P_l - R_{ijkl} \sigma_{ij} \Phi_k \Phi_l \tag{2d}$$

Electrostriction and rotostriction coefficients, $Q_{ijkl}$ and $R_{ijkl}$, are listed in Table I. Elastic and flexoelectric energies are

$$\Delta G_{elast} = -s_{ijkl} \sigma_{ij} \sigma_{kl} - \frac{F_{ijkl}}{2} \left( \sigma_{ij} \frac{\partial P_k}{\partial x_l} - P_k \frac{\partial \sigma_{ij}}{\partial x_l} \right) \tag{2e}$$

Here $s_{ijkl}$ are the components of elastic compliances tensor and $F_{ijkl}$ are flexoelectric tensor components (listed in Table I).

**Table I.** Parameters used in LGD calculations for AFD-FE perovskite BFO

| Parameter | Designation | Numerical value for BFO |
|---|---|---|
| Effective permittivity | $\varepsilon_{eff} = \Sigma_i \varepsilon_{bi} + \varepsilon_{el}$ | 160 |
| dielectric stiffness | $\alpha_T$ (×10$^5$ C$^{-2}$·Jm/K) | 9 |
| Curie temperature for P | $T_C$ (K) | 1300 |
| Barret temperature for P | $T_{qP}$ (K) | 800 |
| polar expansion 4$^{th}$ order | $a_{ij}$ (×10$^8$ C$^{-4}$·m$^5$J) | $a_{11}$= –13.5, $a_{12}$= 5 |
| LGD expansion 6$^{th}$ order | $a_{ijk}$ (×10$^9$ C$^{-6}$·m$^9$J) | $a_{111}$= 11.2, $a_{112}$= –3, $a_{123}$= –6 |
| electrostriction | $Q_{ij}$ (C$^{-2}$·m$^4$) | $Q_{11}$=0.054, $Q_{12}$= –0.015, $Q_{44}$=0.02 |
| Stiffness components | $c_{ij}$ (×10$^{11}$ Pa) | $c_{11}$=3.02, $c_{12}$= 1.62, $c_{44}$=0.68 |
| polarization gradient coefficients | $g_{ij}$ (×10$^{-10}$ C$^{-2}$m$^3$J) | $g_{11}$=10, $g_{12}$= –7, $g_{44}$=5 |
| AFD-FE coupling | $\xi_{ij}$ (×10$^{29}$ C$^{-2}$·m$^{-2}$ J/K) | $\xi_{11}$ = –0.5, $\xi_{12}$ =0.5, $\xi_{44}$ = –2.6 |
| tilt expansion 2$^{nd}$ order | $b_T$ (×10$^{26}$·J/(m$^5$K)) | 4 |
| Curie temperature for Φ | $T_\Phi$ (K) | 1440 |
| Barret temperature for Φ | $T_{q\Phi}$ (K) | 400 |
| tilt expansion 4$^{nd}$ order | $b_{ij}$ (×10$^{48}$ J/(m$^7$)) | $b_{11}$= –24+4.5 $(\coth(300/T) - \coth(3/14))$ $b_{12}$= 45–4.5 $(\coth(300/T) - \coth(1/4))$ |
| tilt expansion 6$^{nd}$ order | $b_{ijk}$ (×10$^{70}$ J/(m$^9$)) | $b_{111}$= 4.5–3.4 $(\coth(400/T) - \coth(2/7))$ $b_{112}$= 3.6–0.04 $(\coth(10/T) - \coth(1/130))$ $b_{123}$= 41–43.2 $(\coth(1200/T) - \coth(12/11))$ |
| tilt gradient coefficients | $v_{ij}$ (×10$^{11}$ J/ m$^3$) | $v_{11}$=2, $v_{12}$=-1, $v_{44}$=1 |
| rotostriction | $R_{ij}$ (×10$^{18}$ m$^{-2}$) | $R_{11}$= –1.32, $R_{12}$= –0.43, $R_{44}$=8.45 |
| Flexoelectric coefficients | $F_{ij}$ (×10$^{-11}$ m$^3$/C) | $F_{11}$= 2, $F_{12}$= 1, $F_{44}$= 0.5 |

The surface energy of the film has the form:

$$\int_S \left( \frac{b_i^{(S)}}{2} \Phi_i^2 + \frac{a_i^{(S)}}{2} P_i^2 \right) dx_1 dx_3 \quad (3)$$

Surface energy coefficients $b_i^{(S)}$ and $a_i^{(S)}$ have different nature and control the broadening at the surface of ADF and FE domain walls, respectively.

The coupled system of Euler-Lagrange equations allowing for Khalatnikov relaxation of the oxygen tilt and polarization components $\Phi_i$ and $P_i$ is:

$$\frac{\delta G}{\delta P_i} = -\Gamma \frac{\partial P_i}{\partial t} \text{ and } \frac{\delta G}{\delta \Phi_i} = -\Gamma \frac{\partial \Phi_i}{\partial t}. \quad (4a)$$

These equations are supplemented by the boundary conditions of zero generalized fluxes at the film boundaries,

$$\left. b_i^{(S)} \Phi_i + v_{ijkl} \frac{\partial \Phi_j}{\partial x_k} n_l \right|_{x_2=0,h} = 0, \quad \left. a_i^{(S)} P_i + g_{ijkl} \frac{\partial P_j}{\partial x_k} n_l \right|_{x_2=0,h} = 0, \quad (4b)$$

without summation on $i$=1, 2, 3 and $n_l$ are the components of external normal to the film surfaces $x_2 = 0, h$. The explicit form of equations (4) is listed in the Appendix B of Suppl. Mat. The natural boundary conditions used hereinafter, correspond to the case $b_i^{(S)} = 0$ and $a_i^{(S)} = 0$.

Elastic problem formulation is based on the equations of state, $u_{ij} = -\frac{\delta G}{\delta \sigma_{kl}}$, where $u_{ij}$ are elastic strain tensor components. Mechanical equilibrium conditions are $\partial \sigma_{ij}/\partial x_j = 0$ [28].

The misfit strain $u_m$ appears from the difference between the lattice constants of the film and substrate [31], and so we suppose that misfit strain, $u_{11} = u_{33} = u_m$, is applied into XZ-plane at the BFO-PSO interface $x_2 = 0$. In order to estimate the misfit strain for BFO film on orthorhombic PSO substrates at room temperature (RT), we used pseudo-cubic lattice constant obtained as cubic root of the volume of elementary cell (see Table II). Note that the misfit strain $u_m$ contributes to the elastic boundary conditions:

$$\left. \sigma_{ij} n_j \right|_{x_2=h} = 0, \quad \left. (U_1 - u_m x_1) \right|_{x_2=0} = 0, \quad \left. U_2 \right|_{x_2=0} = 0, \quad \left. (U_3 - u_m x_3) \right|_{x_2=0} = 0 \quad (5)$$

**Table II.** Parameters used for misfit strain estimation in BFO thin films on PSO substrate

| Parameter description | Numerical value |
| --- | --- |
| Lattice constants of PrScO$_3$ at RT (Å) | $a$ = 5.608; $b$ = 5.780; $c$ = 8.025 |
| pseudo-cubic "lattice constant" (Å) | 4.021 * |
| BFO film thickness $h$ (nm) | 18 nm, 7 nm |
| BFO pseudo-cubic lattice constant (Å) | 3.9916+6.9×10$^{-5}$($T$-295) ** |
| Misfit strain $u_m$ (%) at RT | 1.8 % |

* see Gesing et al.[29]

** The dependence was extrapolated to room temperature from higher temperature using the data of Ref. [30]

**Modeling of polarization and tilt behavior in a thin BiFeO$_3$ film on PrScO$_3$ substrate**

The problem described above [see Eqs.(1)-(5)] was analyzed numerically using FEM for a (5 – 20) nm thick BFO film on PSO substrate. The morphology of nominally uncharged DWs in thin BFO film is shown in Figs. 6–8. The film orientation with respect to substrate was {101} and {110}, respectively. Different orientations of DWs are considered below, namely 180° and 71° walls, which schematics are shown in Fig. 1b-c. The values of the gradient coefficients chosen for Figs.6-8 correspond to the maximal qualitative similarity with experimental results for all films.

Unusual morphologies of the 180° DW structure in a {101}-oriented 18-nm BFO film epitaxially grown on PSO substrate are presented in Figs. 6. Curved DWs appeared to be long-living (meta)stable distributions of polarization and tilt, which have been relaxed at times $t \gg \tau_K$ from the initial ($t$=0) straight orientation of the DW at room temperature. It is seen that all components of tilt change its sign across the incline AFD-FE domain wall. Two triangle-like regions with opposite orientation of the tilt component $\Phi_1$ across curved 180° DWs are presented in Fig. 6a at $t$=13$\tau_K$. The relaxed curved and meandering profiles of $\Phi_2$ are shown in Fig. 6b at $t$=13$\tau_K$ and 26$\tau_K$. Noticeably curved and "zig-zag" like meandering $\Phi_3$ distribution are shown in Fig. 6c at $t$=13$\tau_K$. The tilt walls can meander freely because AFD component of the DW is not affected by any sort of "de-elastification" field.

At the same time the polarization components $P_1$ and $P_2$ are affected by depolarization electric field (produced by $div\vec{P} \neq 0$), but its influence significantly decreases with $\varepsilon_{eff}$ increase above 100. At $\varepsilon_{eff} \geq 100$ $P_1$ is almost zero everywhere, but reveal clearly visibly thin meandering regions at the walls (see Fig. 6d). The profiles of $P_2$ across 180° DWs reveal noticable meandering superimposed on a triangle-like motif (see Fig. 6e). The relative straightening of $P_{1,2}$-walls happens only at $\varepsilon_{eff} \approx \varepsilon_b \leq 10$, because the system tries to minimize the high energy of electric depolarization film that is produced by any sort of charged DWs[27], [32-33], and the only way to do this is to keep $P_1 \approx 0$ and $P_2$-walls relatively straight. The distribution $P_3(x_1,x_2)$ is not affected by depolarization field. Therefore the meandering of $P_3$-walls meandering, which originates from the biquadratic AFD-FE coupling and surface influence via the natural boundary conditions, correlates with meandering $\Phi_{2,3}$-walls (compare Fig. 6f with Fig. 6b,c).

The features of meandering morphologies of the 180° DW structure "survives" in a {101}-oriented 18-nm BFO film even after a very long relaxation time (see Figs. 7). Specifically, curved and meandering DWs appeared thermodynamically stable at room temperature. Several curved regions with nonzero $\Phi_1$ of the opposite orientation, small and evolved meanders at domain walls of $\Phi_2$ and $\Phi_3$ are seen in Figs. 7a-c. The profiles of polarization components $P_1$, $P_2$ and $P_3$ correlate with the meandering of $\Phi_2$ and $\Phi_3$ walls (see Figs. 7d-f), but the depolarization keeps $P_2$-walls more straight.

As it was shown earlier without consideration of misfit effect, electrostrictive and rotostriction coupling[33], and confirmed in this work with these effects included in thin BFO films, the origin of the meandering 180° AFD-FE walls is conditioned by the decrease of the tilts gradient energy. We checked numerically that the origin of the meandering walls does not steam from incomplete polarization screening in thin BFO films, electrostrictive, rotostrictive or flexoelectric coupling. The spatial confinement delineates the appropriate boundary conditions for the oxygen tilt and polarization components at the film surfaces, but its existence is not critical for the appearance of meandering walls and their zig-zag stability. The values of the gradient energy coefficients for the oxygen tilt appeared critical to initiate the morphological changes of the 180° uncharged domain walls towards zig-zag meandering. Meandering instability appears for small gradient energies, while the walls become more straight and broaden at higher gradients.

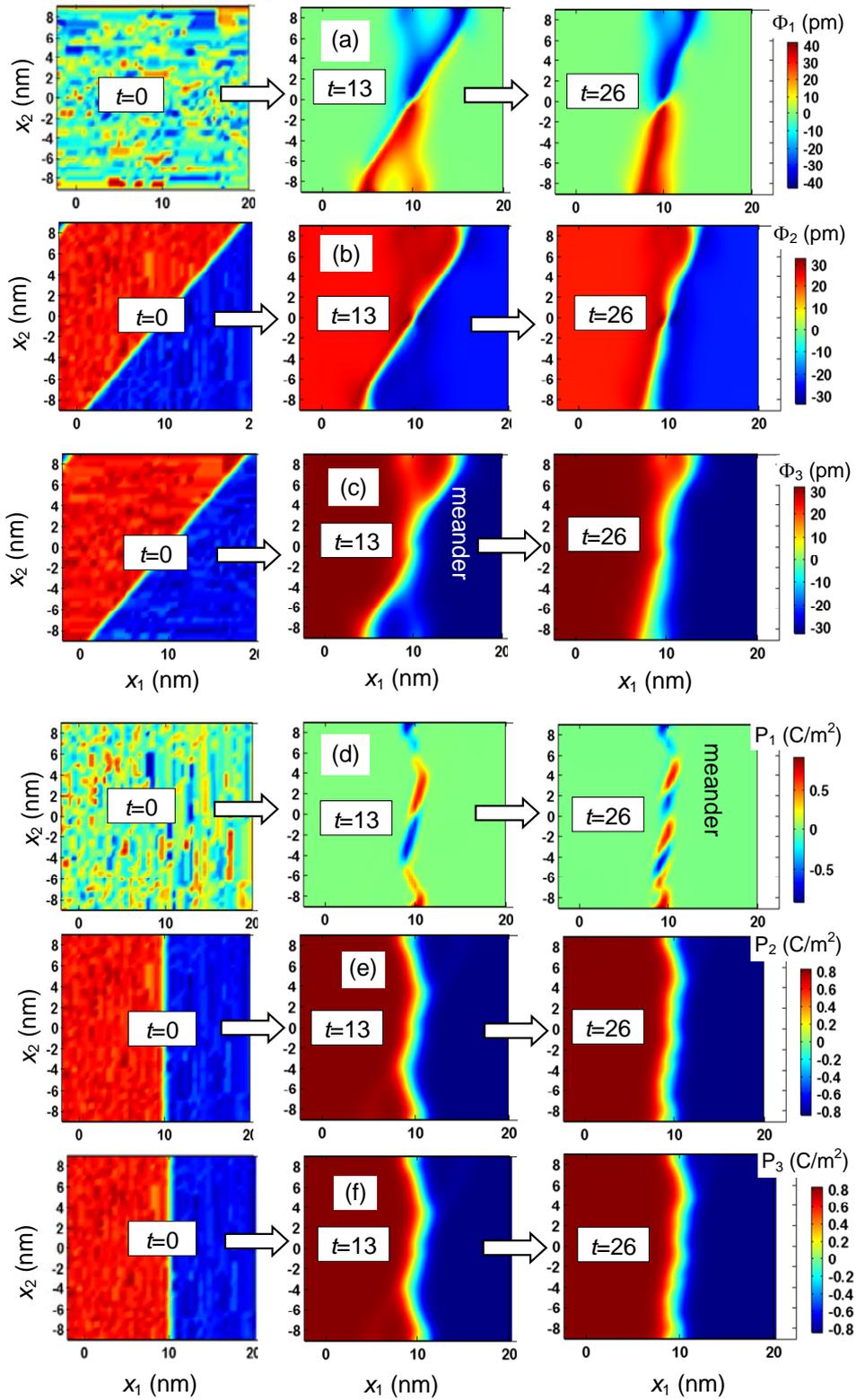

**Figure 6|{101} oriented 180° DWs in 18 nm BFO film on PrScO$_3$ substrate, relaxing from initial straight DWs at $t=0$ to the intermediate state $t=13\tau_K$, and than to the final state $t>25\tau_K$.** Distribution of the AFD order parameter $\Phi_i$ **(a)-(c)** and polarization $P_i$ **(d)-(f)** components in the cross-section of a thin BFO film with {101} oriented 180° DWs. Film thickness is 18 nm, room temperature. Tilted coordinate system is used for $x_1$.

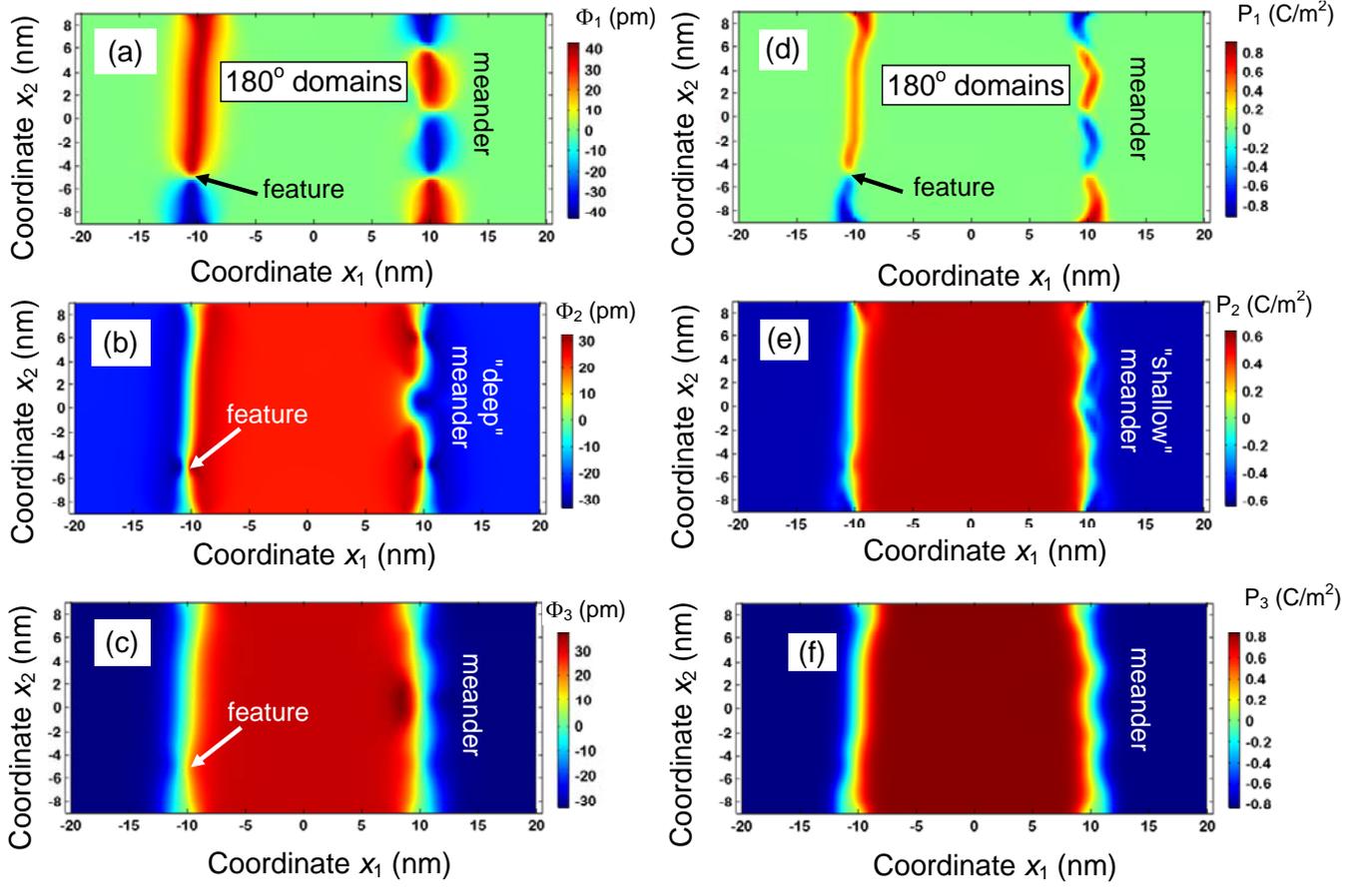

**Figure 7| {101} oriented 180° DWs in 18 nm BFO film on PrScO$_3$ substrate, after a very long relaxation time.** Distribution of the AFD order parameter $\Phi_i$ **(a)-(c)** and polarization $P_i$ **(d)-(f)** components in the $\{x_1 x_2\}$ cross-section of a thin BFO film with {101} oriented 180° DWs. Film thickness is 18 nm, room temperature. Tilted coordinate system is used for $x_1$.

Typical morphologies of the 71° DW structure in a {110}-oriented 7-nm BFO film epitaxially grown on PSO substrate are shown in Figs. 8. Bulk 71° domains correspond to the case, when only one component of vectorial order parameter changes its sign when crossing the wall plane. For the case DWs are inclined and curved near the electrodes for all components of polarization and tilt. No meandering instability or zig-zag changes appear in the case, but the DW width increases with the tilt gradient coefficient increase. The inclination of 71° DW is determined by the conditions of electrical and mechanical compatibility, while the bending of DW near the film surfaces is induced by the surface influence. Namely, since we used the natural boundary conditions, it means that the corresponding gradients of order parameters are zero at the surface, and hence the DWs should be perpendicular to the surface near the surface, while it is inclined in the central part of the film.

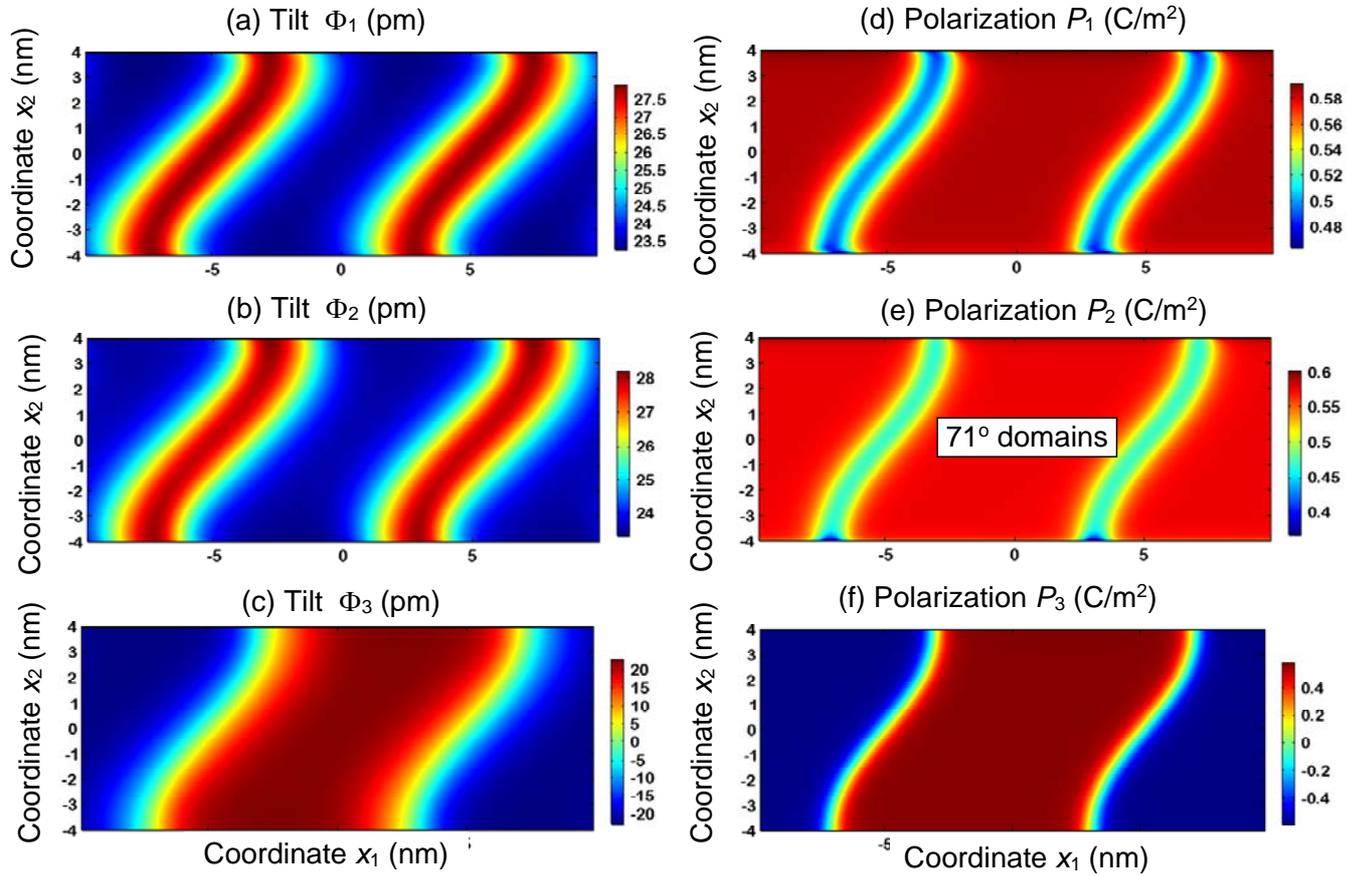

**Figure 8| {110} oriented 71° DWs in BFO film on PrScO$_3$ substrate (fully relaxed).** Distribution of different components of the AFD order parameter $\Phi_i$ (a)-(c) and polarization $P_i$ (d)-(f) in the $\{x_1x_2\}$ cross-section of 7-nm BFO film with {110} oriented 71° DWs, room temperature.

In accordance with theoretical estimates (see Appendix E in Suppl.Mat), the experimentally observed conductivity enhancement at the domain wall can be caused by the electric potential and elastic strain variations inside the wall. The electro-elastic potential relief (wells or humps) leads to higher concentration of electrons (or holes) in the DW due to the local band bending. It appeared that the contribution of the strain variations to the conductivity modulation is dominant, because the potential is zero exactly at the 180° and 71° DWs. Since the strain variation is proportional to $Q_{ijkl}P_kP_l + R_{ijkl}\Phi_k\Phi_l + F_{ijkl}\frac{\partial P_k}{\partial x_l}$, it may also include the atomic bond changes at the wall via the striction and flexoelectric mechanisms. Figure S2 shows the map of normalized static conductivity (that is directly proportional to CAFM contrast) across {101} oriented 180° DWs in 18 nm BFO film on PSO substrate. From the figure the DW can be one order of magnitude more conductive than the domains itself.

**Discussion**

Based on the results above, we perform both experimental and theoretical investigations on the morphology and AFD-FE coupling at 180° and 71° domain walls. The comparison between the experimental and theoretical results give more hints for AFD-FE coupling and corresponding functional manipulation.

Experimental results suggest that there are several kinks at 180° domain walls, and corresponding polarization components also demonstrate some kinks. Despite the fact that the calculated AFD order parameters are larger than the experimental results, which may originates from the uncertain coefficients of coupling of strain with tilt and polarization, the AFD characteristics analysed in experimental and theoretical share a lot common features. The order parameter $\Phi_1$ in Fig. 6a (corresponds to the oxygen displacements analyzed in Fig. 2e) reveals similar two triangle-like regions with opposite displacements for each domain wall. More specifically, for the case the upward polarization in left and downward polarization in right side of the wall, the oxygen columns would end up with a positive shift triangle-like region near the surface and a negative shift triangle-like region near the interface. For the case the downward polarization in left and upward polarization in right side of the wall, the oxygen columns would end up with a negative shift triangle-like region near the surface and a positive shift triangle-like region near the interface. Note that the obvious decrease of oxygen tilt components near the interface observed experimentally is not corroborated by theoretical results for $\Phi_1$ shown in Fig. 6a. This may be related with different boundary conditions between experiments and calculations. The order parameter $\Phi_2$ in Fig. 6b corresponds to the oxygen column displacements along $[001]_{pc}$ direction in Fig. 2f, where the inclination and meandering of AFD-FE walls are obvious. Moreover, theoretical results confirm that the values of the gradient energy coefficients for the oxygen tilt are critical to initiate the morphological changes of the 180° uncharged domain walls towards zig-zag meandering. The meandering instability would appear at AFD-FE walls for small gradient energies, whereas the FE walls would become more straight and broaden at higher gradients (see Fg.S2 in Suppl,Mat)_.

For 71° AFD-FE walls, the experimental and theoretical results are much close to each other, including the broaden AFD walls compared with FE walls, and the curvature of AFD-FE walls at the interface and surface. The meandering instability and triangle-like regions appeared in 180° AFD-FE walls are not observed in 71° AFD-FE walls. LGD theory proves that the width of AFD wall increases with the tilt gradient coefficient, and the inclination of 71° DW is determined by the conditions of electrical and mechanical compatibility, while the bending of DW near the film surfaces is induced by the surface influence.

Based on the analyses above, we can conclude that the tilt gradient energy would greatly influence the AFD-FE coupling at domain walls, and induce peculiar AFD-FE wall morphology according to different kinds of domain walls. Since the oxygen octahedra are reported to be strongly coupled with the electronic, magnetic and optical properties[14,15], we asume that the different morphorlogy induced by tilt gradient changes may corresponds to the specific functionalities according to each DW.

Besides, we observed a significantly enhanced conductivity at the 180° AFD-FE walls. In accordance with theoretical estimates (see Appendix E in Suppl.Mat), the conductivity enhancement at the domain wall can be caused by the electric potential changes and elastic strain variations inside the wall. Further experimental investigations demonstrate an obvious decrease of Fe-O-Fe bond angles at these 180° domain walls, which overturn the previous unstanding that the reduced bandgap is induced by strengthening Fe-O-Fe bond angles at 180° domain walls. On the contrary, we propose that it is the reduction of Fe-O-Fe bond angle that induce high conductivity at uncharged 180° domain walls.

## Conclusion

The coupling of AFD and FE long-range orders at the nominally uncharged 180° and 71° domain walls are investigated both experimentally and theoretically. Quite unexpectedly, 180° AFD-FE walls demonstrate a rather unusual meandering characteristics and oxygen displacements triangle-like regions. However, in 71° domain walls, the AFD-FE coupling would result in broaden AFD walls and an obvious curvature near the interface and surface. LGD theory calculations agree with the experimental results very well and confirmed that the origin of the unusual AFD-FE walls is conditioned by the decrease of the tilts gradient energy in thin BFO films. For 180° AFD-FE walls, meandering instability appears for small gradient energies, while the walls become more straight and broaden at higher gradients[33]. However, uncharged 71° walls in thin films do not reveal any meandering instability, but their width increase with the tilt gradient coefficient increase. Moreover, we observe an obvious conduction at 180° AFD-FE walls accompanied by a decrease of Fe-O-Fe bond angles at 180° domain walls which overturn the previous unstanding that the reduced bandgap is induced by strengthening Fe-O-Fe bond angles at 180° domain walls.

Based on this, we revealed experimentally and explained theoretically the previously unexplored type of the gradient-driven morphological phase transition taking place at the AFD-FE domain walls in thin strained multiferroic films. We also discover an unusual reduction of B-O-B bond angle which may corresponds to the conduction at domain walls. These results help us better understand the AFD-FE coupling at domain walls and explore its possibility to induce physical properties for future nano-devices.

## Methods

**Material preparation.** We used Pulsed Laser Deposition (PLD) to grow 7 nm and 18 nm $BiFeO_3$ (BFO) films on $PrScO_3$ (PSO; $a$ = 5.608 Å; $b$ = 5.780 Å; $c$ = 8.025 Å) $(001)_O$ (subscript 'O' denotes the orthorhombic orientation) substrates. The laser we used is a Coherent ComPex PRO 201 F KrF ($\lambda$=248 nm) excimer laser. The target we used is a sintered stoichiometric BFO ceramic with 1 inch in diameter. Commercial, one-side polishing PSO $(001)_O$ oriented substrates with 10 mm × 10 mm × 0.5 mm dimension were used for film deposition. All of the substrates were heated up to 850°C for 20 minutes before deposition to clean the surface. The target-substrate distance was set at 32 mm. During the deposition, the substrate temperature was set at 800 °C, the oxygen pressure was set at 12 Pa, the repetition rate was set at 6 Hz and the laser energy applied was 2 J·cm$^{-2}$. After deposition, the samples were kept at 800 °C for 5 minutes and then cooled down to room temperature with 5K·min$^{-1}$ in an oxygen pressure of $3\times10^4$ Pa.

**Scanning transmission electron microscopy.** A traditional method (slicing, gluing, grinding, dimpling and ion milling) is used to prepare the specimens for HAADF-STEM imaging. For the final ion milling we used a Gatan 691 PIPS at a voltage lower than 1kV to reduce the ion beam damage. A Titan Cubed 60–300 kV microscope (FEI) was used to acquire the atomic resolved HAADF and ABF images. The aberration-corrected scanning transmission electron microscope is fitted with a monochromator, a high-brightness field-emission gun (X-FEG) and double aberration (Cs) correctors operating at 300 kV. The semi-convergence angle is set at 21.4 mrad. To carry out the atom column positions, we used Matlab software to fit them as 2D Gaussian peaks[34-36].

**Piezoresponse Force Microscopy and Conductive AFM characterizations.** The local piezoresponse was carried out by PFM on a commercial AFM system (Cypher, Asylum Research) in ambient conditions at room temperature. The contact frequency for VPFM measurements is set around 350 kHz. The conductive AFM (CAFM) was carried out by the same system with an Orca holder using Ti/Ir (5/20) coated tips (ASYELEC-01-R) with nominal $k$ = 2.8 N·m$^{-1}$. To perform CAFM on the cross-sectional sample, a thin graphite layer was deposited on the backside of the specimen.

**Data availability.** The data that support the findings of this study are available from the corresponding author upon reasonable request.

**Acknowledgements.** This work is supported by the Key Research Program of Frontier Sciences CAS (QYZDJ-SSW-JSC010), the National Natural Science Foundation of China (No. 51671194, No.51571197), and National Basic Research Program of China (2014CB921002). Y. L. T. acknowledges the IMR SYNL-T.S. Kê Research Fellowship and the Youth Innovation Promotion Association CAS (No. 2016177). We are grateful to Mr. B. Wu and Mr. L.X. Yang of this lab for their technical support on the Titan platform of G2 60-300kV aberration-corrected scanning transmission electron microscope. A.N.M work has received funding from the European Union's Horizon 2020 research and innovation programme under the Marie Skłodowska-Curie grant agreement No 778070, and partially supported by the National Academy of Sciences of Ukraine (project No. 0117U002612).

**Authors' contribution.** X.L.M. and Y.L.Z. conceived the project of interfacial characterization in oxides by using aberration-corrected STEM. M.J.H., Y.L.Z., Y.L.T. and X.L.M. designed the experiments. M.J.H. performed the thin-film growth, STEM observations, and PFM/AFM measurement. Y.J.W. and X.W.G carried out digital analysis of the STEM data. E.A.E. wrote the codes, performed numerical calculations and prepared figures. A.N.M. generated the research idea, stated the problem, analyzed theoretical results and wrote the theory part of the manuscript. All authors contributed to the discussions and manuscript preparation.

**Competing interests:** The authors declare no competing interests.

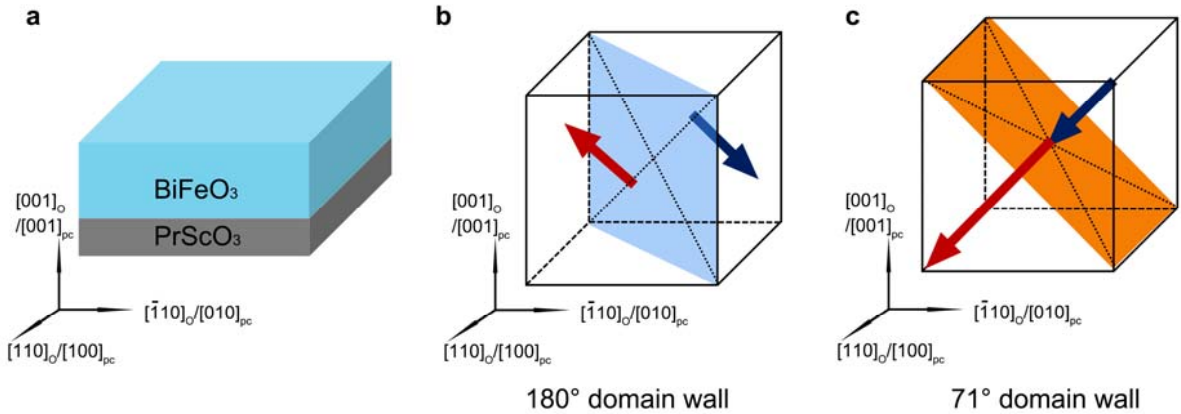

**Figure S1| Schematics showing the experiment system.** BFO/PSO $(001)_O$ interface (a), 180° domain walls (b) and 71° domain walls (c) according to experimental results.

# APPENDIX A

## Evident form of the free energy of antiferrodistortive ferroelectric bulk

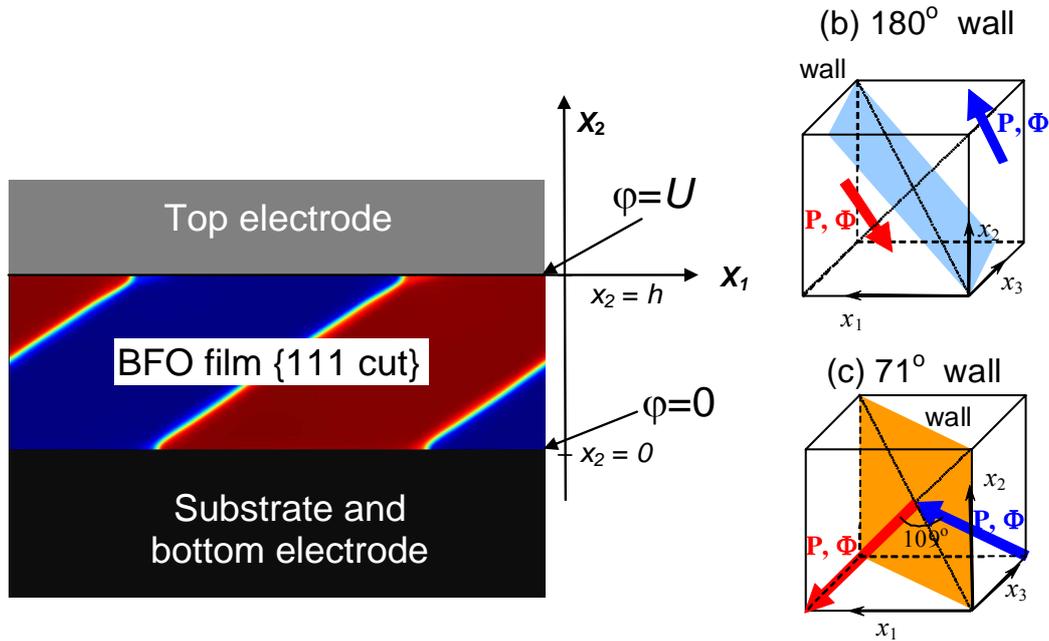

**Figure S2| (a)** Considered system, consisting of thin BiFeO$_3$ (BFO) film, {111}-cut with a domain wall (DW) structure, placed between electrically conducting top and bottom electrodes. Two types of the considered nominally uncharged 180- **(b)** and 71-degree **(c)** domain walls in BFO are shown in the right column. BFO/PSO pair corresponds to the mismatch strain $u_m = +1.8\%$.

We use phenomenological Landau-Ginsburg-Devonshire (LGD) theory to model the domain walls (DW) morphology in a thin BFO film placed on a thick rigid substrate. Two vectorial order parameters, namely polarization components $P_i$ and oxygen octahedral tilts $\Phi_i$ are considered for the description of ferroelectric (FE) and antiferrodistortive (AFD) components of the DWs ($i=1, 2, 3$). The bulk part of LGD thermodynamic potential consists of the following contributions:

$$G_V = \int_S dx_1 dx_3 \int_0^h (\Delta G_{AFD} + \Delta G_{FE} + \Delta G_{BQC} + \Delta G_{striction} + \Delta G_{elast} + \Delta G_{flexo}) dx_2 \qquad (A.1)$$

In Eq.(A.1) all contributions are consistent with the parent phase m3m symmetry in accordance with the basics of LGD approach. The compact and explicit forms of the AFD contribution are:

$$\begin{aligned}
\Delta G_{AFD} &= b_i(T)\Phi_i^2 + b_{ij}\Phi_i^2\Phi_j^2 + b_{ijk}\Phi_i^2\Phi_j^2\Phi_k^2 + v_{ijkl}\frac{\partial \Phi_i}{\partial x_k}\frac{\partial \Phi_j}{\partial x_l} \\
&\equiv b_1(\Phi_1^2 + \Phi_2^2 + \Phi_3^2) + b_{11}(\Phi_1^4 + \Phi_2^4 + \Phi_3^4) + b_{12}(\Phi_1^2\Phi_2^2 + \Phi_1^2\Phi_3^2 + \Phi_3^2\Phi_2^2) + \\
&\quad b_{111}(\Phi_1^6 + \Phi_2^6 + \Phi_3^6) + b_{112}(\Phi_1^2(\Phi_2^4 + \Phi_3^4) + \Phi_1^4(\Phi_2^2 + \Phi_3^2) + \Phi_2^2\Phi_3^4 + \Phi_2^4\Phi_3^2) + b_{123}\Phi_1^2\Phi_2^2\Phi_3^2 + \\
&\quad + \frac{v_{11}}{2}\left(\left(\frac{\partial \Phi_1}{\partial x_1}\right)^2 + \left(\frac{\partial \Phi_2}{\partial x_2}\right)^2 + \left(\frac{\partial \Phi_3}{\partial x_3}\right)^2\right) + v_{12}\left(\frac{\partial \Phi_1}{\partial x_1}\frac{\partial \Phi_2}{\partial x_2} + \frac{\partial \Phi_2}{\partial x_2}\frac{\partial \Phi_3}{\partial x_3} + \frac{\partial \Phi_1}{\partial x_1}\frac{\partial \Phi_3}{\partial x_3}\right) + \\
&\quad \frac{v_{44}}{2}\left(\left(\frac{\partial \Phi_1}{\partial x_2}\right)^2 + \left(\frac{\partial \Phi_1}{\partial x_3}\right)^2 + \left(\frac{\partial \Phi_2}{\partial x_1}\right)^2 + \left(\frac{\partial \Phi_2}{\partial x_3}\right)^2 + \left(\frac{\partial \Phi_3}{\partial x_1}\right)^2 + \left(\frac{\partial \Phi_3}{\partial x_2}\right)^2\right) + \\
&\quad + v'_{44}\left(\frac{\partial \Phi_1}{\partial x_2}\frac{\partial \Phi_2}{\partial x_1} + \frac{\partial \Phi_1}{\partial x_3}\frac{\partial \Phi_3}{\partial x_1} + \frac{\partial \Phi_2}{\partial x_3}\frac{\partial \Phi_3}{\partial x_2}\right)
\end{aligned} \qquad (A.2)$$

In accordance with the classical LGD theory, we assume that the coefficients $b_i$ are temperature dependent in accordance with a Barrett law [i], $b_i = b_T T_{q\Phi}(\coth(T_{q\Phi}/T) - \coth(T_{q\Phi}/T_\Phi))$, where $T_\Phi$ is the AFD transition temperature and $T_{q\Phi}$ is a characteristic temperature [ii]. Description and numerical values of the phenomenological coefficients $b_i$, $b_{ij}$, $b_{ijk}$ and gradient coefficients $v_{ij}$ included in Eq.(A.2) can be found in **Table I**, where Voight notations are used.

The compact and explicit forms of the FE contribution are:

$$\begin{aligned}
\Delta G_{FE} &= a_i P_i^2 + a_{ij}P_i^2 P_j^2 + a_{ijk}P_i^2 P_j^2 P_k^2 + g_{ijkl}\frac{\partial P_i}{\partial x_k}\frac{\partial P_j}{\partial x_l} - P_i E_i \\
&= a_1(P_1^2 + P_2^2 + P_3^2) + a_{11}(P_1^4 + P_2^4 + P_3^4) + a_{12}(P_1^2 P_2^2 + P_1^2 P_3^2 + P_2^2 P_3^2) \\
&\quad + a_{111}(P_1^6 + P_2^6 + P_3^6) + a_{112}(P_1^2(P_2^4 + P_3^4) + P_1^4(P_2^2 + P_3^2) + P_2^2 P_3^4 + P_2^4 P_3^2) + a_{123}P_1^2 P_2^2 P_3^2 \\
&\quad + \frac{g_{11}}{2}\left(\left(\frac{\partial P_1}{\partial x_1}\right)^2 + \left(\frac{\partial P_2}{\partial x_2}\right)^2 + \left(\frac{\partial P_3}{\partial x_3}\right)^2\right) + g_{12}\left(\frac{\partial P_1}{\partial x_1}\frac{\partial P_2}{\partial x_2} + \frac{\partial P_2}{\partial x_2}\frac{\partial P_3}{\partial x_3} + \frac{\partial P_1}{\partial x_1}\frac{\partial P_3}{\partial x_3}\right) + \\
&\quad + \frac{g_{44}}{2}\left(\left(\frac{\partial P_1}{\partial x_2}\right)^2 + \left(\frac{\partial P_1}{\partial x_3}\right)^2 + \left(\frac{\partial P_2}{\partial x_1}\right)^2 + \left(\frac{\partial P_2}{\partial x_3}\right)^2 + \left(\frac{\partial P_3}{\partial x_1}\right)^2 + \left(\frac{\partial P_3}{\partial x_2}\right)^2\right) + \\
&\quad + g'_{44}\left(\frac{\partial P_1}{\partial x_2}\frac{\partial P_2}{\partial x_1} + \frac{\partial P_1}{\partial x_3}\frac{\partial P_3}{\partial x_1} + \frac{\partial P_2}{\partial x_3}\frac{\partial P_3}{\partial x_2}\right) - P_i E_i
\end{aligned} \qquad (A.3)$$

In accordance with the LGD theory, we assume that the coefficients $a_k$ are temperature dependent and obeys the Barrett law, $a_k^{(P)} = \alpha_T(T_{qP}\coth(T_{qP}/T) - T_C)$, where $T_C$ is the Curie temperature and $T_{qP}$ is a characteristic temperature [i, ii]. Description and numerical values of the phenomenological coefficients

$a_i$, $a_{ij}$, $a_{ijk}$ and gradient coefficients $g_{ij}$ included in Eq.(A.3) can be found in **Table I**, and the Voight notations are used.

Electric field components are $E_i$, which are defined by electrostatic potential in the conventional way, $E_i = -\partial\varphi/\partial x_i$. The potential satisfies Poisson equation, $\varepsilon_0 \varepsilon_{\mathit{eff}} \dfrac{\partial^2 \varphi}{\partial x_i^2} = \dfrac{\partial P_i}{\partial x_i}$, that is supplied by electric boundary conditions ($\varepsilon_b$ is a background permittivity [iii]). Electric boundary conditions for the short-circuited film are the zero electric potential at the conducting electrodes, $\varphi|_{x_2=0,h} = 0$.

The compact and explicit forms of the biquadratic coupling between polarization and tilt is

$$\Delta G_{BQC} = \zeta_{ijkl} \Phi_i \Phi_j P_k P_l$$
$$= \zeta_{11}(\Phi_1^2 P_1^2 + \Phi_2^2 P_2^2 + \Phi_3^2 P_3^2) + \zeta_{12}((\Phi_2^2 + \Phi_3^2)P_1^2 + (\Phi_1^2 + \Phi_3^2)P_2^2 + (\Phi_1^2 + \Phi_2^2)P_3^2), \quad (A.4)$$
$$+ \zeta_{44}(\Phi_1 \Phi_2 P_1 P_2 + \Phi_1 \Phi_3 P_1 P_3 + \Phi_2 \Phi_3 P_2 P_3)$$

As one can see, the coupling energy (A.4) includes poorly known tensorial coefficients AFD-FE biquadratic couplings $\zeta_{44}$, $\zeta_{11}$ and $\zeta_{12}$, which have been treated as fitting parameters to experiment (listed in **Table I**).

Electrostriction and rotostriction contributions are

$$\Delta G_{striction} = -Q_{ijkl}\sigma_{ij}P_k P_l - R_{ijkl}\sigma_{ij}\Phi_k \Phi_l = -Q_{11}(\sigma_{11}P_1^2 + \sigma_{22}P_2^2 + \sigma_{33}P_3^2)$$
$$- Q_{12}((\sigma_{22} + \sigma_{33})P_1^2 + (\sigma_{11} + \sigma_{33})P_2^2 + (\sigma_{22} + \sigma_{11})P_3^2)$$
$$- Q_{44}(\sigma_{12}P_1 P_2 + \sigma_{13}P_1 P_3 + \sigma_{23}P_2 P_3) \quad (A.5)$$
$$- R_{11}(\sigma_{11}\Phi_1^2 + \sigma_{22}\Phi_2^2 + \sigma_{33}\Phi_3^2) - R_{12}((\sigma_{22} + \sigma_{33})\Phi_1^2 + (\sigma_{11} + \sigma_{33})\Phi_2^2 + (\sigma_{22} + \sigma_{11})\Phi_3^2)$$
$$- R_{44}(\sigma_{12}\Phi_1 \Phi_2 + \Phi_1 \Phi_3 +_{23} \Phi_2 \Phi_3)$$

Electrostriction and rotostriction coefficients, $Q_{ijkl}$ and $R_{ijkl}$, are listed in **Table I**. Elastic energy is

$$\Delta G_{elast} = -s_{ijkl}\sigma_{ij}\sigma_{kl}$$
$$= -\frac{s_{11}}{2}(\sigma_{11}^2 + \sigma_{22}^2 + \sigma_{33}^2) - s_{12}(\sigma_{11}\sigma_{22} + \sigma_{22}\sigma_{33} + \sigma_{11}\sigma_{33}) - \frac{s_{44}}{2}(\sigma_{12}^2 + \sigma_{23}^2 + \sigma_{13}^2) \quad (A.6)$$

Flexoelectric effect contribution has the form

$$\Delta G_{\mathrm{flexo}} = -\frac{F_{ijkl}}{2}\left(\sigma_{ij}\frac{\partial P_k}{\partial x_l} - P_k \frac{\partial \sigma_{ij}}{\partial x_l}\right)$$
$$= -F_{11}\left(\sigma_{11}\frac{\partial P_1}{\partial x_1} + \sigma_{22}\frac{\partial P_2}{\partial x_2} + \sigma_{33}\frac{\partial P_3}{\partial x_3}\right)$$
$$- F_{12}\left((\sigma_{22} + \sigma_{33})\frac{\partial P_1}{\partial x_1} + (\sigma_{11} + \sigma_{33})\frac{\partial P_2}{\partial x_2} + (\sigma_{11} + \sigma_{22})\frac{\partial P_3}{\partial x_3}\right) \quad (A.7)$$
$$- F_{44}\left(\sigma_{12}\left(\frac{\partial P_1}{\partial x_2} + \frac{\partial P_2}{\partial x_1}\right) + \sigma_{13}\left(\frac{\partial P_1}{\partial x_3} + \frac{\partial P_3}{\partial x_1}\right) + \sigma_{23}\left(\frac{\partial P_2}{\partial x_3} + \frac{\partial P_3}{\partial x_2}\right)\right)$$

$F_{ijkl}$ are flexoelectric tensor components, which values are listed in **Table I** (their typical range $0 \leq |F_{ijkl}| \leq 10^{11}$ m$^3$/C).

**Table I.** Parameters used in LGD calculations for "tilted" perovskite BFO

| Parameter | Designation | Numerical value for BFO |
|---|---|---|
| Effective permittivity including background, Jahn-Teller modes and electronic contributions | $\varepsilon_{eff} = \Sigma_i \varepsilon_{bi} + \varepsilon_{el}$ | 160 |
| dielectric stiffness | $\alpha_T$ ($\times 10^5$ C$^{-2}\cdot$Jm/K) | 9 |
| Curie temperature for P | $T_C$ (K) | 1300 |
| Barret temperature for P | $T_{qP}$ (K) | 800 |
| polar expansion 4$^{th}$ order | $a_{ij}$ ($\times 10^8$ C$^{-4}\cdot$m$^5$J) | $a_{11} = -13.5$, $a_{12} = 5$ |
| LGD expansion 6$^{th}$ order | $a_{ijk}$ ($\times 10^9$ C$^{-6}\cdot$m$^9$J) | $a_{111} = 11.2$, $a_{112} = -3$, $a_{123} = -6$ |
| electrostriction | $Q_{ij}$ (C$^{-2}\cdot$m$^4$) | $Q_{11} = 0.054$, $Q_{12} = -0.015$, $Q_{44} = 0.02$ |
| Stiffness components | $c_{ij}$ ($\times 10^{11}$ Pa) | $c_{11} = 3.02$, $c_{12} = 1.62$, $c_{44} = 0.68$ |
| polarization gradient coefficients | $g_{ij}$ ($\times 10^{-10}$ C$^{-2}$m$^3$J) | $g_{11} = 10$, $g_{12} = -7$, $g_{44} = 5$ |
| AFD-FE coupling | $\times 10^{29}$ C$^{-2}\cdot$m$^{-2}$ J/K | $\xi_{11} = -0.5$, $\xi_{12} = 0.5$, $\xi_{44} = -2.6$ |
| tilt expansion 2$^{nd}$ order | $b_T$ ($\times 10^{26}\cdot$J/(m$^5$K)) | 4 |
| Curie temperature for $\Phi$ | $T_\Phi$ (K) | 1440 |
| Barret temperature for $\Phi$ | $T_{q\Phi}$ (K) | 400 |
| tilt expansion 4$^{nd}$ order | $b_{ij}$ ($\times 10^{48}$ J/(m$^7$)) | $b_{11} = -24 + 4.5 \left( \coth(300/T) - \coth(3/14) \right)$<br>$b_{12} = 45 - 4.5 \left( \coth(300/T) - \coth(1/4) \right)$ |
| tilt expansion 6$^{nd}$ order | $b_{ijk}$ ($\times 10^{70}$ J/(m$^9$)) | $b_{111} = 4.5 - 3.4 \left( \coth(400/T) - \coth(2/7) \right)$<br>$b_{112} = 3.6 - 0.04 \left( \coth(10/T) - \coth(1/130) \right)$<br>$b_{123} = 41 - 43.2 \left( \coth(1200/T) - \coth(12/11) \right)$ |
| tilt gradient coefficients | $v_{ij}$ ($\times 10^{11}$ J/m$^3$) | $v_{11} = 2$, $v_{12} = -1$, $v_{44} = 1$ |
| rotostriction | $R_{ij}$ ($\times 10^{18}$ m$^{-2}$) | $R_{11} = -1.32$, $R_{12} = -0.43$, $R_{44} = 8.45$ |
| Flexoelectric coefficients | $F_{ij}$ ($\times 10^{-11}$ m$^3$/C) | $F_{11} = 2$, $F_{12} = 1$, $F_{44} = 0.5$ |

## APPENDIX B.

### Euler-Lagrange equations with boundary conditions

The Euler-Lagrange equations of LGD theory for polarization $P_i$ and tilt $\Phi_i$ components ($i = 1, 2, 3$) could be obtained by the minimization of the free energy (A.1) as follows

$$
\begin{aligned}
& 2P_1 \left( a_1 - Q_{12}(\sigma_{22} + \sigma_{33}) - Q_{11}\sigma_{11} \right) - Q_{44}(\sigma_{12}P_2 + \sigma_{13}P_3) + \left( \zeta_{11}\Phi_1^2 + \zeta_{12}(\Phi_2^2 + \Phi_3^2) \right) 2P_1 + \zeta_{44}\Phi_1(\Phi_2 P_2 + \Phi_3 P_3) \\
& + 4a_{11}P_1^3 + 2a_{12}P_1(P_2^2 + P_3^2) + 6a_{111}P_1^5 + 2a_{112}P_1(P_2^4 + 2P_1^2 P_2^2 + P_3^4 + 2P_1^2 P_3^2) + 2a_{112}P_1 P_2^2 P_3^2 \\
& - g_{11}\frac{\partial^2 P_1}{\partial x_1^2} - g_{44}\left( \frac{\partial^2 P_1}{\partial x_2^2} + \frac{\partial^2 P_1}{\partial x_3^2} \right) - (g'_{44} + g_{12})\frac{\partial^2 P_2}{\partial x_2 \partial x_1} - (g'_{44} + g_{12})\frac{\partial^2 P_3}{\partial x_3 \partial x_1} \\
& + F_{11}\frac{\partial \sigma_{11}}{\partial x_1} + F_{12}\left( \frac{\partial \sigma_{22}}{\partial x_1} + \frac{\partial \sigma_{33}}{\partial x_1} \right) + F_{44}\left( \frac{\partial \sigma_{12}}{\partial x_2} + \frac{\partial \sigma_{13}}{\partial x_3} \right) = E_1
\end{aligned}
\quad \text{(B.1a)}
$$

$$2P_2(a_1 - Q_{12}(\sigma_{11} + \sigma_{33}) - Q_{11}\sigma_{22}) - Q_{44}(\sigma_{12}P_1 + \sigma_{23}P_3) + (\zeta_{11}\Phi_2^2 + \zeta_{12}(\Phi_1^2 + \Phi_3^2))2P_2 + \zeta_{44}\Phi_2(\Phi_1 P_1 + \Phi_3 P_3)$$
$$+ 4a_{11}P_2^3 + 2a_{12}P_2(P_1^2 + P_3^2) + 6a_{111}P_2^5 + 2a_{112}P_2(P_1^4 + 2P_2^2 P_1^2 + P_3^4 + 2P_2^2 P_3^2) + 2a_{112}P_2 P_1^2 P_3^2$$
$$- g_{11}\frac{\partial^2 P_2}{\partial x_2^2} - g_{44}\left(\frac{\partial^2 P_2}{\partial x_1^2} + \frac{\partial^2 P_2}{\partial x_3^2}\right) - (g'_{44} + g_{12})\frac{\partial^2 P_1}{\partial x_2 \partial x_1} - (g'_{44} + g_{12})\frac{\partial^2 P_3}{\partial x_3 \partial x_2} \quad \text{(B.1b)}$$
$$+ F_{11}\frac{\partial \sigma_{22}}{\partial x_2} + F_{12}\left(\frac{\partial \sigma_{11}}{\partial x_2} + \frac{\partial \sigma_{33}}{\partial x_2}\right) + F_{44}\left(\frac{\partial \sigma_{12}}{\partial x_1} + \frac{\partial \sigma_{23}}{\partial x_3}\right) = E_2$$

$$2P_3(a_1 - Q_{12}(\sigma_{11} + \sigma_{22}) - Q_{11}\sigma_{33}) - Q_{44}(\sigma_{13}P_1 + \sigma_{23}P_2) + (\zeta_{11}\Phi_3^2 + \zeta_{12}(\Phi_1^2 + \Phi_2^2))2P_3 + \zeta_{44}\Phi_3(\Phi_1 P_1 + \Phi_2 P_2)$$
$$+ 4a_{11}P_3^3 + 2a_{12}P_3(P_1^2 + P_2^2) + 6a_{111}P_3^5 + 2a_{112}P_3(P_1^4 + 2P_3^2 P_1^2 + P_2^4 + 2P_2^2 P_3^2) + 2a_{112}P_3 P_1^2 P_2^2$$
$$- g_{11}\frac{\partial^2 P_3}{\partial x_3^2} - g_{44}\left(\frac{\partial^2 P_3}{\partial x_1^2} + \frac{\partial^2 P_3}{\partial x_2^2}\right) - (g'_{44} + g_{12})\frac{\partial^2 P_1}{\partial x_3 \partial x_1} - (g'_{44} + g_{12})\frac{\partial^2 P_2}{\partial x_3 \partial x_2} \quad \text{(A.8c)}$$
$$+ F_{11}\frac{\partial \sigma_{33}}{\partial x_3} + F_{12}\left(\frac{\partial \sigma_{11}}{\partial x_3} + \frac{\partial \sigma_{33}}{\partial x_3}\right) + F_{44}\left(\frac{\partial \sigma_{13}}{\partial x_1} + \frac{\partial \sigma_{23}}{\partial x_2}\right) = E_3$$

$$(b_1 - R_{12}(\sigma_{22} + \sigma_{33}) - R_{11}\sigma_{11} + \zeta_{11}P_1^2 + \zeta_{12}(P_2^2 + P_3^2))2\Phi_1 - R_{44}(\sigma_{12}\Phi_2 + \sigma_{13}\Phi_3) + \zeta_{44}P_1(\Phi_2 P_2 + \Phi_3 P_3)$$
$$+ 4b_{11}\Phi_1^3 + 2b_{12}\Phi_1(\Phi_2^2 + \Phi_3^2) + 6b_{111}\Phi_1^5 + 2b_{112}\Phi_1(\Phi_2^4 + 2\Phi_1^2\Phi_2^2 + \Phi_3^4 + 2\Phi_1^2\Phi_3^2) + 2b_{112}\Phi_1\Phi_2^2\Phi_3^2 \quad \text{(B.2a)}$$
$$- \nu_{11}\frac{\partial^2 \Phi_1}{\partial x_1^2} - \nu_{44}\left(\frac{\partial^2 \Phi_1}{\partial x_2^2} + \frac{\partial^2 \Phi_1}{\partial x_3^2}\right) - (\nu'_{44} + \nu_{12})\frac{\partial^2 \Phi_2}{\partial x_1 \partial x_2} - (\nu'_{44} + \nu_{12})\frac{\partial^2 \Phi_3}{\partial x_1 \partial x_3} = 0$$

$$(b_1 - R_{12}(\sigma_{11} + \sigma_{33}) - R_{11}\sigma_{22} + \zeta_{11}P_2^2 + \zeta_{12}(P_1^2 + P_3^2))2\Phi_2 - R_{44}(\sigma_{12}\Phi_1 + \sigma_{23}\Phi_3) + \zeta_{44}P_2(\Phi_1 P_1 + \Phi_3 P_3)$$
$$+ 4b_{11}\Phi_2^3 + 2b_{12}\Phi_2(\Phi_1^2 + \Phi_3^2) + 6b_{111}\Phi_2^5 + 2b_{112}\Phi_2(\Phi_1^4 + 2\Phi_1^2\Phi_2^2 + \Phi_3^4 + 2\Phi_2^2\Phi_3^2) + 2b_{112}\Phi_2\Phi_1^2\Phi_3^2 \quad \text{(B.2b)}$$
$$- \nu_{11}\frac{\partial^2 \Phi_2}{\partial x_2^2} - \nu_{44}\left(\frac{\partial^2 \Phi_2}{\partial x_1^2} + \frac{\partial^2 \Phi_2}{\partial x_3^2}\right) - (\nu'_{44} + \nu_{12})\frac{\partial^2 \Phi_1}{\partial x_1 \partial x_2} - (\nu'_{44} + \nu_{12})\frac{\partial^2 \Phi_3}{\partial x_2 \partial x_3} = 0$$

$$(b_1 - R_{12}(\sigma_{22} + \sigma_{11}) - R_{11}\sigma_{33} + \zeta_{11}P_3^2 + \zeta_{12}(P_2^2 + P_1^2))2\Phi_3 - R_{44}(\sigma_{23}\Phi_2 + \sigma_{13}\Phi_1) + \zeta_{44}P_3(\Phi_2 P_2 + \Phi_1 P_1)$$
$$+ 4b_{11}\Phi_3^3 + 2b_{12}\Phi_3(\Phi_1^2 + \Phi_2^2) + 6b_{111}\Phi_3^5 + 2b_{112}\Phi_3(\Phi_2^4 + 2\Phi_3^2\Phi_2^2 + \Phi_1^4 + 2\Phi_1^2\Phi_3^2) + 2b_{112}\Phi_3\Phi_1^2\Phi_2^2 \quad \text{(B.2c)}$$
$$- \nu_{11}\frac{\partial^2 \Phi_3}{\partial x_3^2} - \nu_{44}\left(\frac{\partial^2 \Phi_3}{\partial x_2^2} + \frac{\partial^2 \Phi_3}{\partial x_1^2}\right) - (\nu'_{44} + \nu_{12})\frac{\partial^2 \Phi_1}{\partial x_1 \partial x_3} - (\nu'_{44} + \nu_{12})\frac{\partial^2 \Phi_2}{\partial x_2 \partial x_3} = 0$$

The boundary conditions are

$$b^{(S)}\Phi_i + \nu_{ijkl}\frac{\partial \Phi_k}{\partial x_l}n_j\bigg|_S = 0, \quad a^{(P)}P_i + \left(g_{ijkl}\frac{\partial P_k}{\partial x_l} - F_{klij}\sigma_{kl}\right)n_j\bigg|_S = 0 \quad (i=1, 2, 3) \quad \text{(B.3)}$$

For the normal $n_l = \{0, \pm 1, 0\}$ to the surface $x_2 = 0, h$ have the following explicit form

$$b^{(S)}\Phi_1 \mp \left(\nu_{44}\frac{\partial \Phi_1}{\partial x_2} + \nu'_{44}\frac{\partial \Phi_2}{\partial x_1}\right)\bigg|_{x_2=0,h} = 0, \quad b^{(S)}\Phi_3 \mp \left(\nu_{44}\frac{\partial \Phi_3}{\partial x_2}\right)\bigg|_{x_2=0,h} = 0, \quad \text{(B.4a)}$$

$$b^{(S)}\Phi_2 \mp \left(\nu_{11}\frac{\partial \Phi_2}{\partial x_2} + \nu_{12}\frac{\partial \Phi_1}{\partial x_1}\right)\bigg|_{x_2=0,h} = 0 \quad \text{(B.4b)}$$

$$a^{(S)}P_1 \mp \left(g_{44}\frac{\partial P_1}{\partial x_2} + g'_{44}\frac{\partial P_2}{\partial x_1} - F_{44}\sigma_{12}\right)\bigg|_{x_2=0,h} = 0, \quad a^{(S)}P_3 \mp \left(g_{44}\frac{\partial P_3}{\partial x_2} - F_{44}\sigma_{23}\right)\bigg|_{x_2=0,h} = 0,$$

$$a^{(S)}P_2 \mp \left(g_{11}\frac{\partial P_2}{\partial x_2} + g_{12}\frac{\partial P_1}{\partial x_1} - F_{11}\sigma_{22} - F_{12}(\sigma_{11}+\sigma_{33})\right)\bigg|_{x_2=0,h} = 0 \quad (B.5)$$

Note that all derivatives $\frac{\partial}{\partial x_3}$ are zero for the 2D-problem. Note that following Glinka and Marton semi-microscopic model, we suggested that

$$g'_{44} + g_{12} \equiv 0 \text{ and } v'_{44} + v_{12} \equiv 0 \quad (B.6)$$

Elastic stresses satisfy the equation of mechanical equilibrium in the film, $\frac{\partial \sigma_{ij}}{\partial x_j} = 0$. Equations of state (modified Hooke's law) should be obtained from the variation of the energy with respect to elastic stress, $\frac{\delta G}{\delta \sigma_{ij}} = -u_{ij}$, namely:

$$u_{11} = s_{11}\sigma_{11} + s_{12}(\sigma_{22}+\sigma_{33}) + Q_{11}P_1^2 + Q_{12}(P_2^2+P_3^2) + R_{11}\Phi_1^2 + R_{12}(\Phi_2^2+\Phi_3^2) + F_{11}\frac{\partial P_1}{\partial x_1} + F_{12}\left(\frac{\partial P_2}{\partial x_2} + \frac{\partial P_3}{\partial x_3}\right)$$

$$u_{22} = s_{12}(\sigma_{11}+\sigma_{33}) + s_{11}\sigma_{22} + Q_{12}(P_1^2+P_3^2) + Q_{11}P_2^2 + R_{12}(\Phi_1^2+\Phi_3^2) + R_{11}\Phi_2^2 + F_{11}\frac{\partial P_2}{\partial x_2} + F_{12}\left(\frac{\partial P_1}{\partial x_1} + \frac{\partial P_3}{\partial x_3}\right)$$

$$u_{33} = s_{12}(\sigma_{11}+\sigma_{22}) + s_{11}\sigma_{33} + Q_{12}(P_1^2+P_2^2) + Q_{11}P_3^2 + R_{12}(\Phi_1^2+\Phi_2^2) + R_{11}\Phi_3^2 + F_{11}\frac{\partial P_3}{\partial x_3} + F_{12}\left(\frac{\partial P_2}{\partial x_2} + \frac{\partial P_1}{\partial x_1}\right)$$

$$u_{12} = \frac{1}{2}\left(s_{44}\sigma_{12} + Q_{44}P_1P_2 + R_{44}\Phi_1\Phi_2 + F_{44}\left(\frac{\partial P_1}{\partial x_2} + \frac{\partial P_2}{\partial \hat{x}_1}\right)\right)$$

$$u_{13} = \frac{1}{2}\left(s_{44}\sigma_{12} + Q_{44}P_1P_3 + R_{44}\Phi_1\Phi_3 + F_{44}\left(\frac{\partial P_1}{\partial x_3} + \frac{\partial P_3}{\partial x_1}\right)\right)$$

$$u_{23} = \frac{1}{2}\left(s_{44}\sigma_{23} + Q_{44}P_3P_2 + R_{44}\Phi_3\Phi_2 + F_{44}\left(\frac{\partial P_2}{\partial x_3} + \frac{\partial P_3}{\partial x_2}\right)\right)$$

(B.7)

The misfit strain stems from the difference between the lattice constants of film and substrate and so we suppose that misfit strain, $u_{11} = u_{33} = u_m$, is applied into XZ-plane at the film surface $x_2 = 0$. Corresponding elastic boundary conditions are:

$$\sigma_{12}|_{x_2=h} = 0, \quad \sigma_{22}|_{x_2=h} = 0, \quad \sigma_{23}|_{x_2=h} = 0, \quad (B.8a)$$

$$(U_1 - u_m x_1)|_{x_2=0} = 0, \quad U_2|_{x_2=0} = 0, \quad (U_3 - u_m x_3)|_{x_2=0} = 0 \quad (B.8b)$$

# APPENDIX C.

## Rotated frame for the [101] domain walls consideration

In many important cases ferroic domain walls are inclined with respect of the axes of the coordinate system related to aristo-phase (which is cubic one in the case of perovskites). It is natural to suppose that the physical fields depend only on the distances to the walls and the film surface for simplest pseudo-1D domain structures in films/slabs (i.e. when all the domain walls are perpendicular to selected plane). Therefore, it is convenient to perform calculations in the frame where one of the coordinate axis (denoted by $X_3$) is parallel to domain walls and to the developed surface of the film. Then one could suppose that the considered problem is independent on coordinate $x_3$.

Let us consider the system of coordinates, rotated on 45 degrees from initial ones around $X_2$ axis. The components of vectors and tensors in this "rotated frame" will be denoted with "cap", namely, one have the following relations between the components in the initial frame and the rotated one:

$$\hat{x}_1 = \frac{x_1 - x_3}{\sqrt{2}}, \quad \hat{x}_2 = x_2, \quad \hat{x}_3 = \frac{x_1 + x_3}{\sqrt{2}}. \tag{C.1a}$$

$$\hat{P}_1 = \frac{P_1 - P_3}{\sqrt{2}}, \quad \hat{P}_2 = P_2, \quad \hat{P}_3 = \frac{P_1 + P_3}{\sqrt{2}}. \tag{C.1b}$$

$$\hat{\Phi}_1 = \frac{\Phi_1 - \Phi_3}{\sqrt{2}}, \quad \hat{\Phi}_2 = \Phi_2, \quad \hat{\Phi}_3 = \frac{\Phi_1 + \Phi_3}{\sqrt{2}}. \tag{C.1c}$$

$$\hat{\sigma}_{11} = \frac{\sigma_{11} + \sigma_{33}}{2} - \sigma_{13}, \quad \hat{\sigma}_{12} = \frac{\sigma_{12} - \sigma_{23}}{\sqrt{2}}, \quad \hat{\sigma}_{13} = \frac{\sigma_{11} - \sigma_{33}}{2},$$

$$\hat{\sigma}_{22} = \sigma_{22}, \quad \hat{\sigma}_{23} = \frac{\sigma_{12} + \sigma_{23}}{\sqrt{2}}, \quad \hat{\sigma}_{33} = \frac{\sigma_{11} + \sigma_{33}}{2} + \sigma_{13}. \tag{C.1d}$$

Free energy contribution of antiferrodistortive phase (Eq. (A.2)) could be written via vectors (C.1) as follows:

$$\Delta G_{AFD} = b_1\left(\hat{\Phi}_1^2 + \Phi_2^2 + \hat{\Phi}_3^2\right) + \hat{b}_{11}\left(\hat{\Phi}_1^4 + \hat{\Phi}_3^4\right) + b_{11}\Phi_2^4 + \hat{b}_{13}\hat{\Phi}_1^2\hat{\Phi}_3^2 + b_{12}\left(\hat{\Phi}_1^2\Phi_2^2 + \hat{\Phi}_3^2\Phi_2^2\right) +$$
$$b_{111}\Phi_2^6 + \hat{b}_{111}\left(\hat{\Phi}_1^6 + \hat{\Phi}_3^6\right) + b_{112}\Phi_2^4\left(\hat{\Phi}_1^2 + \hat{\Phi}_3^2\right) + \hat{b}_{112}\left(\hat{\Phi}_1^4\Phi_2^2 + \hat{\Phi}_3^4\Phi_2^2\right) + \hat{b}_{113}\left(\hat{\Phi}_3^4\hat{\Phi}_1^2 + \hat{\Phi}_1^4\hat{\Phi}_3^2\right) + \hat{b}_{123}\hat{\Phi}_1^2\Phi_2^2\hat{\Phi}_3^2 +$$
$$\frac{\hat{v}_{11}}{2}\left(\left(\frac{\partial\hat{\Phi}_1}{\partial\hat{x}_1}\right)^2 + \left(\frac{\partial\hat{\Phi}_3}{\partial\hat{x}_3}\right)^2\right) + \hat{v}_{13}\frac{\partial\hat{\Phi}_1}{\partial\hat{x}_1}\frac{\partial\hat{\Phi}_3}{\partial\hat{x}_3} + v_{12}\left(\frac{\partial\hat{\Phi}_1}{\partial\hat{x}_1}\frac{\partial\Phi_2}{\partial x_2} + \frac{\partial\Phi_2}{\partial x_2}\frac{\partial\hat{\Phi}_3}{\partial\hat{x}_3}\right) + \frac{\hat{v}_{55}}{2}\left(\left(\frac{\partial\hat{\Phi}_1}{\partial\hat{x}_3}\right)^2 + \left(\frac{\partial\hat{\Phi}_3}{\partial\hat{x}_1}\right)^2\right) +$$
$$\frac{v_{11}}{2}\left(\frac{\partial\Phi_2}{\partial x_2}\right)^2 + \frac{v_{44}}{2}\left(\left(\frac{\partial\hat{\Phi}_1}{\partial x_2}\right)^2 + \left(\frac{\partial\hat{\Phi}_3}{\partial x_2}\right)^2 + \left(\frac{\partial\Phi_2}{\partial\hat{x}_1}\right)^2 + \left(\frac{\partial\Phi_2}{\partial\hat{x}_3}\right)^2\right) + v'_{44}\left(\frac{\partial\hat{\Phi}_1}{\partial x_2}\frac{\partial\Phi_2}{\partial\hat{x}_1} + \frac{\partial\Phi_2}{\partial\hat{x}_3}\frac{\partial\hat{\Phi}_3}{\partial x_2}\right) + \hat{v}'_{55}\frac{\partial\hat{\Phi}_1}{\partial\hat{x}_3}\frac{\partial\hat{\Phi}_3}{\partial\hat{x}_1}$$

(C.2)

Here we introduced the components of tensors in the rotated frame

$$\hat{b}_{11} = \frac{b_{11}}{2} + \frac{b_{12}}{4}, \quad \hat{b}_{13} = 3b_{11} - \frac{b_{12}}{2}; \tag{C.3a}$$

$$\hat{b}_{111} = \frac{b_{111}}{4} + \frac{b_{112}}{4}, \ \hat{b}_{112} = \frac{b_{112}}{2} + \frac{b_{123}}{4}, \ \hat{b}_{113} = \frac{15 b_{111}}{4} - \frac{b_{112}}{4}, \ \hat{b}_{123} = 3 b_{112} - \frac{b_{123}}{2};  \quad \text{(C.3b)}$$

$$\hat{v}_{11} = \frac{v_{11} + v_{12} + v_{44} + v'_{44}}{2}, \ \hat{v}_{13} = \frac{v_{11} + v_{12} - v_{44} - v'_{44}}{2}. \quad \text{(C.3c)}$$

$$\hat{v}_{55} = \frac{v_{11} - v_{12} + v_{44} - v'_{44}}{2}, \ \hat{v}'_{55} = \frac{v_{11} - v_{12} - v_{44} + v'_{44}}{2}, \quad \text{(C.3d)}$$

Contribution of ferroelectric order parameter is (compare with (A.3)):

$$\begin{aligned}
\Delta G_{FE} &= a_1\left(\hat{P}_1^2 + P_2^2 + \hat{P}_3^2\right) + \hat{a}_{11}\left(\hat{P}_1^4 + \hat{P}_3^4\right) + a_{11} P_2^4 + \hat{a}_{13} \hat{P}_1^2 \hat{P}_3^2 + a_{12}\left(\hat{P}_1^2 P_2^2 + P_2^2 \hat{P}_3^2\right) \\
&+ \hat{a}_{111}\left(\hat{P}_1^6 + \hat{P}_3^6\right) + a_{111} P_2^6 + \hat{a}_{112}\left(\hat{P}_1^4 P_2^2 + \hat{P}_3^4 P_2^2\right) + a_{112} P_2^4\left(\hat{P}_1^2 + \hat{P}_3^2\right) + \hat{a}_{113}\left(\hat{P}_3^4 \hat{P}_1^2 + \hat{P}_1^4 \hat{P}_3^2\right) + \hat{a}_{123} \hat{P}_1^2 P_2^2 \hat{P}_3^2 \\
&+ \frac{\hat{g}_{11}}{2}\left(\left(\frac{\partial \hat{P}_1}{\partial \hat{x}_1}\right)^2 + \left(\frac{\partial \hat{P}_3}{\partial \hat{x}_3}\right)^2\right) + \frac{g_{11}}{2}\left(\frac{\partial P_2}{\partial x_2}\right)^2 + \hat{g}_{13} \frac{\partial \hat{P}_1}{\partial \hat{x}_1} \frac{\partial \hat{P}_3}{\partial \hat{x}_3} + \hat{g}'_{55} \frac{\partial \hat{P}_1}{\partial \hat{x}_3} \frac{\partial \hat{P}_3}{\partial \hat{x}_1} + \frac{\hat{g}_{55}}{2}\left(\left(\frac{\partial \hat{P}_1}{\partial \hat{x}_3}\right)^2 + \left(\frac{\partial \hat{P}_3}{\partial \hat{x}_1}\right)^2\right) \\
&+ g_{12}\left(\frac{\partial \hat{P}_1}{\partial \hat{x}_1} \frac{\partial P_2}{\partial x_2} + \frac{\partial P_2}{\partial x_2} \frac{\partial \hat{P}_3}{\partial \hat{x}_3}\right) + g'_{44}\left(\frac{\partial \hat{P}_1}{\partial x_2} \frac{\partial P_2}{\partial \hat{x}_1} + \frac{\partial P_2}{\partial \hat{x}_3} \frac{\partial \hat{P}_3}{\partial x_2}\right) + \frac{g_{44}}{2}\left(\left(\frac{\partial \hat{P}_1}{\partial x_2}\right)^2 + \left(\frac{\partial \hat{P}_3}{\partial x_2}\right)^2 + \left(\frac{\partial P_2}{\partial \hat{x}_1}\right)^2 + \left(\frac{\partial P_2}{\partial \hat{x}_3}\right)^2\right)
\end{aligned}$$

(C.4)

Here we introduced the components of tensors in the rotated frame

$$\hat{a}_{11} = \frac{a_{11}}{2} + \frac{a_{12}}{4}, \quad \hat{a}_{13} = 3 a_{11} - \frac{a_{12}}{2}; \quad \text{(C.5a)}$$

$$\hat{a}_{111} = \frac{a_{111}}{4} + \frac{a_{112}}{4}, \ \hat{a}_{112} = \frac{a_{112}}{2} + \frac{a_{123}}{4}, \ \hat{a}_{113} = \frac{15 a_{111}}{4} - \frac{a_{112}}{4}, \ \hat{a}_{123} = 3 a_{112} - \frac{a_{123}}{2}; \quad \text{(C.5b)}$$

$$\hat{g}_{11} = \frac{g_{11} + g_{12} + g_{44} + g'_{44}}{2}, \ \hat{g}_{13} = \frac{g_{11} + g_{12} - g_{44} - g'_{44}}{2}, \quad \text{(C.5c)}$$

$$\hat{g}_{55} = \frac{g_{11} - g_{12} + g_{44} - g'_{44}}{2}, \ \hat{g}'_{55} = \frac{g_{11} - g_{12} - g_{44} + g'_{44}}{2}. \quad \text{(C.5d)}$$

Finally, the coupling between polarization and tilt (see Eq.(A.4)) in the rotated frame is

$$\begin{aligned}
\Delta G_{BQC} &= \hat{\zeta}_{11}\left(\hat{P}_1^2 \hat{\Phi}_1^2 + \hat{P}_3^2 \hat{\Phi}_3^2\right) + \zeta_{11} P_2^2 \Phi_2^2 + \zeta_{12}\left(\left(\hat{P}_1^2 + \hat{P}_3^2\right)\Phi_2^2 + P_2^2\left(\hat{\Phi}_1^2 + \hat{\Phi}_3^2\right)\right) + \hat{\zeta}_{13}\left(\hat{P}_1^2 \hat{\Phi}_3^2 + \hat{P}_3^2 \hat{\Phi}_1^2\right) \\
&+ \zeta_{44}\left(\hat{P}_1 P_2 \hat{\Phi}_1 \Phi_2 + P_2 \hat{P}_3 \Phi_2 \hat{\Phi}_3\right) + \hat{\zeta}_{55} \hat{P}_1 \hat{P}_3 \hat{\Phi}_1 \hat{\Phi}_3
\end{aligned}$$

(C.6)

where rotated frame tensor components for the coupling energy (C.6) is introduced as:

$$\hat{\zeta}_{11} = \frac{\zeta_{11}}{2} + \frac{\zeta_{12}}{2} + \frac{\zeta_{44}}{4}, \quad \hat{\zeta}_{13} = \frac{\zeta_{11}}{2} + \frac{\zeta_{12}}{2} - \frac{\zeta_{44}}{4}, \quad \hat{\zeta}_{55} = 2\left(\zeta_{11} - \zeta_{12}\right) \quad \text{(C.7)}$$

Electrostriction and rotostriction contributions are

$$\begin{aligned}
\Delta G_{striction} &= -\hat{Q}_{11}\left(\hat{\sigma}_{11} \hat{P}_1^2 + \hat{\sigma}_{33} \hat{P}_3^2\right) - Q_{11} \sigma_{22} P_2^2 - Q_{12}\left(\sigma_{22}\left(\hat{P}_1^2 + \hat{P}_3^2\right) + \left(\hat{\sigma}_{11} + \hat{\sigma}_{33}\right) P_2^2\right) - \hat{Q}_{13}\left(\hat{\sigma}_{11} \hat{P}_3^2 + \hat{\sigma}_{33} \hat{P}_1^2\right) \\
&- Q_{44}\left(\hat{\sigma}_{12} \hat{P}_1 P_2 + \hat{\sigma}_{23} P_2 \hat{P}_3\right) - \hat{Q}_{55} \hat{\sigma}_{13} \hat{P}_1 \hat{P}_3 \\
&- \hat{R}_{11}\left(\hat{\sigma}_{11} \hat{\Phi}_1^2 + \hat{\sigma}_{33} \hat{\Phi}_3^2\right) - R_{11} \sigma_{22} \Phi_2^2 - R_{12}\left(\sigma_{22}\left(\hat{\Phi}_3^2 + \hat{\Phi}_1^2\right) + \left(\hat{\sigma}_{11} + \hat{\sigma}_{33}\right)\Phi_2^2\right) - \hat{R}_{13}\left(\hat{\sigma}_{11} \hat{\Phi}_3^2 + \hat{\sigma}_{33} \hat{\Phi}_1^2\right) \\
&- R_{44}\left(\hat{\sigma}_{12} \hat{\Phi}_1 \Phi_2 + \hat{\sigma}_{23} \Phi_2 \hat{\Phi}_3\right) - \hat{R}_{55} \hat{\sigma}_{13} \hat{\Phi}_1 \hat{\Phi}_3
\end{aligned}$$

(C.8)

Here we introduced the components of tensors in the rotated frame

$$\hat{Q}_{11} = \frac{Q_{11}+Q_{12}}{2}+\frac{Q_{44}}{4} \equiv \hat{Q}_{33}, \quad \hat{Q}_{13} = \frac{Q_{11}+Q_{12}}{2}-\frac{Q_{44}}{4}, \quad \hat{Q}_{55} = 2(Q_{11}-Q_{12}), \quad \text{(C.9a)}$$

$$\hat{R}_{11} = \frac{R_{11}+R_{12}}{2}+\frac{R_{44}}{4} \equiv \hat{R}_{33}, \quad \hat{R}_{13} = \frac{R_{11}+R_{12}}{2}-\frac{R_{44}}{4}, \quad \hat{R}_{55} = 2(R_{11}-R_{12}), \quad \text{(C.9b)}$$

Elastic energy is

$$\Delta G_{elast} = -\frac{\hat{s}_{11}}{2}(\hat{\sigma}_{11}^2+\hat{\sigma}_{33}^2)-\frac{s_{11}}{2}\sigma_{22}^2-s_{12}(\hat{\sigma}_{11}\sigma_{22}+\sigma_{22}\hat{\sigma}_{33})-\hat{s}_{13}\hat{\sigma}_{11}\hat{\sigma}_{33}-\frac{s_{44}}{2}(\hat{\sigma}_{12}^2+\hat{\sigma}_{23}^2)-\frac{\hat{s}_{55}}{2}\hat{\sigma}_{13}^2 \quad \text{(C.10)}$$

With the following designations for tensor components

$$\hat{s}_{11} = \frac{s_{11}+s_{12}}{2}+\frac{s_{44}}{4} \equiv \hat{s}_{33}, \quad \hat{s}_{13} = \frac{s_{11}+s_{12}}{2}-\frac{s_{44}}{4}, \quad \hat{s}_{55} = 2(s_{11}-s_{12}), \quad \text{(C.11)}$$

Flexoelectric coupling contribution to energy is

$$\Delta G_{flexo} = -\hat{F}_{11}\left(\hat{\sigma}_{11}\frac{\partial \hat{P}_1}{\partial \hat{x}_1}+\hat{\sigma}_{33}\frac{\partial \hat{P}_3}{\partial \hat{x}_3}\right)-F_{11}\sigma_{22}\frac{\partial P_2}{\partial x_2}$$

$$-F_{12}\left(\sigma_{22}\left(\frac{\partial \hat{P}_1}{\partial \hat{x}_1}+\frac{\partial \hat{P}_3}{\partial \hat{x}_3}\right)+(\hat{\sigma}_{11}+\hat{\sigma}_{33})\frac{\partial P_2}{\partial x_2}\right)-\hat{F}_{13}\left(\hat{\sigma}_{11}\frac{\partial \hat{P}_3}{\partial \hat{x}_3}+\hat{\sigma}_{33}\frac{\partial \hat{P}_1}{\partial \hat{x}_1}\right) \quad \text{(C.12)}$$

$$-F_{44}\left(\hat{\sigma}_{12}\left(\frac{\partial \hat{P}_1}{\partial x_2}+\frac{\partial P_2}{\partial \hat{x}_1}\right)+\hat{\sigma}_{23}\left(\frac{\partial P_2}{\partial \hat{x}_3}+\frac{\partial \hat{P}_3}{\partial x_2}\right)\right)-\hat{F}_{55}\hat{\sigma}_{13}\left(\frac{\partial \hat{P}_1}{\partial \hat{x}_3}+\frac{\partial \hat{P}_3}{\partial \hat{x}_1}\right)$$

With the following designations for tensor components

$$\hat{F}_{11} = \frac{F_{11}+F_{12}+F_{44}}{2}, \quad \hat{F}_{13} = \frac{F_{11}+F_{12}-F_{44}}{2}, \quad \hat{F}_{55} = F_{11}-F_{12} \quad \text{(C.13)}$$

The Euler-Lagrange equations of LGD theory for polarization $P_i$ and tilt $\Phi_i$ components ($i$=1, 2, 3) could be obtained by the minimization of the free energy (C.2), (C.4), (C.6), (C.8) and (C.12) as follows

$$2\left(a_1-\hat{Q}_{11}\hat{\sigma}_{11}-Q_{12}\sigma_{22}-\hat{Q}_{13}\hat{\sigma}_{33}+a_{12}P_2^2+\hat{a}_{13}\hat{P}_3^2+\hat{\zeta}_{11}\hat{\Phi}_1^2+\zeta_{12}\Phi_2^2+\hat{\zeta}_{13}\hat{\Phi}_3^2\right)\hat{P}_1+$$

$$\left(\zeta_{44}\hat{\Phi}_1\Phi_2-Q_{44}\hat{\sigma}_{12}\right)P_2+\left(\hat{\zeta}_{55}\hat{\Phi}_1\hat{\Phi}_3-\hat{Q}_{55}\hat{\sigma}_{13}\right)\hat{P}_3+$$

$$4\hat{a}_{11}\hat{P}_1^3+6\hat{a}_{111}\hat{P}_1^5+2a_{112}\hat{P}_1P_2^4+4\hat{a}_{112}\hat{P}_1^3P_2^2+2\hat{a}_{113}\hat{P}_1\hat{P}_3^2(\hat{P}_3^2+2\hat{P}_1^2)+2\hat{a}_{123}\hat{P}_1P_2^2\hat{P}_3^2 \quad \text{(C.14a)}$$

$$-\frac{\partial}{\partial \hat{x}_1}\left(\hat{g}_{11}\frac{\partial \hat{P}_1}{\partial \hat{x}_1}+g_{12}\frac{\partial P_2}{\partial x_2}+\hat{g}_{13}\frac{\partial \hat{P}_3}{\partial \hat{x}_3}-\hat{F}_{11}\hat{\sigma}_{11}-F_{12}\sigma_{22}-\hat{F}_{13}\hat{\sigma}_{33}\right)$$

$$-\frac{\partial}{\partial x_2}\left(g_{44}\frac{\partial \hat{P}_1}{\partial x_2}+g'_{44}\frac{\partial P_2}{\partial \hat{x}_1}-F_{44}\hat{\sigma}_{12}\right)-\frac{\partial}{\partial \hat{x}_3}\left(\hat{g}_{55}\frac{\partial \hat{P}_1}{\partial \hat{x}_3}+\hat{g}'_{55}\frac{\partial \hat{P}_3}{\partial \hat{x}_1}-\hat{F}_{55}\hat{\sigma}_{13}\right)=\hat{E}_1$$

$$2\left(a_1 - Q_{12}\hat{\sigma}_{11} - Q_{11}\sigma_{22} - Q_{12}\hat{\sigma}_{33} + a_{12}P_1^2 + a_{12}\hat{P}_3^2 + \zeta_{12}\hat{\Phi}_1^2 + \zeta_{11}\Phi_2^2 + \zeta_{12}\hat{\Phi}_3^2\right)P_2$$
$$+ \left(\zeta_{44}\hat{\Phi}_1\Phi_2 - Q_{44}\hat{\sigma}_{12}\right)\hat{P}_1 + \left(\zeta_{44}\Phi_2\hat{\Phi}_3 - Q_{44}\hat{\sigma}_{23}\right)\hat{P}_3$$
$$+ 4a_{11}P_2^3 + 6a_{111}P_2^5 + 4a_{112}\left(\hat{P}_1^2 + \hat{P}_3^2\right)P_2^3 + 2\hat{a}_{112}\left(\hat{P}_1^4 + \hat{P}_3^4\right)P_2 + 2\hat{a}_{123}P_2\hat{P}_1^2\hat{P}_3^2 \quad \text{(C.14b)}$$
$$- \frac{\partial}{\partial x_2}\left(g_{11}\frac{\partial P_2}{\partial x_2} + g_{12}\frac{\partial \hat{P}_1}{\partial \hat{x}_1} + g_{12}\frac{\partial \hat{P}_3}{\partial \hat{x}_3} - F_{12}\hat{\sigma}_{11} - F_{11}\sigma_{22} - F_{12}\hat{\sigma}_{33}\right)$$
$$- \frac{\partial}{\partial \hat{x}_1}\left(g_{44}\frac{\partial P_2}{\partial \hat{x}_1} + g'_{44}\frac{\partial \hat{P}_1}{\partial x_2} - F_{44}\hat{\sigma}_{12}\right) - \frac{\partial}{\partial \hat{x}_3}\left(g_{44}\frac{\partial P_2}{\partial \hat{x}_3} + g'_{44}\frac{\partial \hat{P}_3}{\partial x_2} - F_{44}\hat{\sigma}_{23}\right) = \hat{E}_2$$

$$2\left(a_1 - \hat{Q}_{11}\hat{\sigma}_{33} - Q_{12}\sigma_{22} - \hat{Q}_{13}\hat{\sigma}_{11} + a_{12}P_2^2 + \hat{a}_{13}\hat{P}_1^2 + \hat{\zeta}_{11}\hat{\Phi}_3^2 + \zeta_{12}\Phi_2^2 + \hat{\zeta}_{13}\hat{\Phi}_1^2\right)\hat{P}_3$$
$$+ \left(\hat{\zeta}_{55}\hat{\Phi}_1\hat{\Phi}_3 - \hat{Q}_{55}\hat{\sigma}_{13}\right)\hat{P}_1 + \left(\zeta_{44}\Phi_2\hat{\Phi}_3 - Q_{44}\hat{\sigma}_{23}\right)P_2$$
$$+ 4\hat{a}_{11}\hat{P}_3^3 + 6\hat{a}_{111}\hat{P}_3^5 + 2a_{112}\hat{P}_3 P_2^4 + 4\hat{a}_{112}\hat{P}_3^3 P_2^2 + 2\hat{a}_{113}\hat{P}_3\hat{P}_1^2\left(\hat{P}_1^2 + 2\hat{P}_3^2\right) + 2\hat{a}_{123}\hat{P}_3 P_2^2 \hat{P}_1^2 \quad \text{(C.14c)}$$
$$- \frac{\partial}{\partial \hat{x}_3}\left(\hat{g}_{11}\frac{\partial \hat{P}_3}{\partial \hat{x}_3} + g_{12}\frac{\partial P_2}{\partial x_2} + \hat{g}_{13}\frac{\partial \hat{P}_1}{\partial \hat{x}_1} - \hat{F}_{11}\hat{\sigma}_{33} - F_{12}\sigma_{22} - \hat{F}_{13}\hat{\sigma}_{11}\right)$$
$$- \frac{\partial}{\partial x_2}\left(g_{44}\frac{\partial \hat{P}_3}{\partial x_2} + g'_{44}\frac{\partial P_2}{\partial \hat{x}_3} - F_{44}\hat{\sigma}_{23}\right) - \frac{\partial}{\partial \hat{x}_1}\left(\hat{g}_{55}\frac{\partial \hat{P}_3}{\partial \hat{x}_1} + \hat{g}'_{55}\frac{\partial \hat{P}_1}{\partial \hat{x}_3} - \hat{F}_{55}\hat{\sigma}_{13}\right) = \hat{E}_3$$

$$2\left(b_1 - \hat{R}_{11}\hat{\sigma}_{11} - R_{12}\sigma_{22} - \hat{R}_{13}\hat{\sigma}_{33} + b_{12}\Phi_2^2 + b_{13}\hat{\Phi}_3^2 + \hat{\zeta}_{11}\hat{P}_1^2 + \zeta_{12}P_2^2 + \hat{\zeta}_{13}\hat{P}_3^2\right)\hat{\Phi}_1 - R_{44}\hat{\sigma}_{12}\Phi_2 - \hat{R}_{55}\hat{\sigma}_{13}\hat{\Phi}_3$$
$$+ \left(\hat{\zeta}_{55}\hat{\Phi}_3\hat{P}_3 + \zeta_{44}\Phi_2 P_2\right)\hat{P}_1$$
$$+ 4\hat{b}_{11}\hat{\Phi}_1^3 + 6\hat{b}_{111}\hat{\Phi}_1^5 + 2b_{112}\hat{\Phi}_1\Phi_2^4 + 4\hat{b}_{112}\hat{\Phi}_1^3\Phi_2^2 + 2\hat{b}_{113}\hat{\Phi}_1\hat{\Phi}_3^2\left(\hat{\Phi}_3^2 + 2\hat{\Phi}_1^2\right) + 2\hat{b}_{123}\hat{\Phi}_1\Phi_2^2\hat{\Phi}_3^2 \quad \text{(C.15a)}$$
$$- \frac{\partial}{\partial \hat{x}_1}\left(\hat{v}_{11}\frac{\partial \hat{\Phi}_1}{\partial \hat{x}_1} + v_{12}\frac{\partial \Phi_2}{\partial x_2} + \hat{v}_{13}\frac{\partial \hat{\Phi}_3}{\partial \hat{x}_3}\right) - \frac{\partial}{\partial x_2}\left(v_{44}\frac{\partial \hat{\Phi}_1}{\partial x_2} + v'_{44}\frac{\partial \Phi_2}{\partial \hat{x}_1}\right) - \frac{\partial}{\partial \hat{x}_3}\left(\hat{v}_{55}\frac{\partial \hat{\Phi}_1}{\partial \hat{x}_3} + \hat{v}'_{55}\frac{\partial \hat{\Phi}_3}{\partial \hat{x}_1}\right) = 0$$

$$2\left(b_1 - R_{12}\hat{\sigma}_{11} - R_{11}\sigma_{22} - R_{12}\hat{\sigma}_{33} + b_{12}\hat{\Phi}_1^2 + b_{12}\hat{\Phi}_3^2 + \zeta_{12}\hat{P}_1^2 + \zeta_{11}P_2^2 + \zeta_{12}\hat{P}_3^2\right)\Phi_2 - R_{44}\hat{\sigma}_{12}\hat{\Phi}_1 - R_{44}\hat{\sigma}_{13}\hat{\Phi}_3$$
$$+ \zeta_{44}\left(\hat{\Phi}_3\hat{P}_3 + \hat{\Phi}_1\hat{P}_1\right)P_2$$
$$+ 4b_{11}\Phi_2^3 + 6b_{111}\Phi_2^5 + 4b_{112}\left(\hat{\Phi}_1^2 + \hat{\Phi}_3^2\right)\Phi_2^3 + 2\hat{b}_{112}\left(\hat{\Phi}_1^4 + \hat{\Phi}_3^4\right)\Phi_2 + 2\hat{b}_{123}\Phi_2\hat{\Phi}_1^2\hat{\Phi}_3^2 \quad \text{(C.15b)}$$
$$- \frac{\partial}{\partial \hat{x}_1}\left(v_{44}\frac{\partial \Phi_2}{\partial \hat{x}_1} + v'_{44}\frac{\partial \hat{\Phi}_1}{\partial x_2}\right) - \frac{\partial}{\partial x_2}\left(v_{11}\frac{\partial \Phi_2}{\partial x_2} + v_{12}\frac{\partial \hat{\Phi}_1}{\partial \hat{x}_1} + v_{12}\frac{\partial \hat{\Phi}_3}{\partial \hat{x}_3}\right) - \frac{\partial}{\partial \hat{x}_3}\left(v_{44}\frac{\partial \Phi_2}{\partial \hat{x}_3} + v'_{44}\frac{\partial \hat{\Phi}_3}{\partial x_2}\right) = 0$$

$$2\left(b_1 - \hat{R}_{11}\hat{\sigma}_{33} - R_{12}\sigma_{22} - \hat{R}_{13}\hat{\sigma}_{11} + b_{12}\Phi_2^2 + b_{13}\hat{\Phi}_1^2 + \hat{\zeta}_{11}\hat{P}_3^2 + \zeta_{12}P_2^2 + \hat{\zeta}_{13}\hat{P}_1^2\right)\hat{\Phi}_3 - R_{44}\hat{\sigma}_{12}\Phi_2 - \hat{R}_{55}\hat{\sigma}_{13}\hat{\Phi}_1$$
$$+ \left(\hat{\zeta}_{55}\hat{\Phi}_1\hat{P}_1 + \zeta_{44}\Phi_2 P_2\right)\hat{P}_3$$
$$+ 4\hat{b}_{11}\hat{\Phi}_3^3 + 6\hat{b}_{111}\hat{\Phi}_3^5 + 2b_{112}\hat{\Phi}_3\Phi_2^4 + 4\hat{b}_{112}\hat{\Phi}_3^3\Phi_2^2 + 2\hat{b}_{113}\hat{\Phi}_3\hat{\Phi}_1^2\left(\hat{\Phi}_1^2 + 2\hat{\Phi}_3^2\right) + 2\hat{b}_{123}\hat{\Phi}_3\Phi_2^2\hat{\Phi}_1^2 \quad \text{(C.15c)}$$
$$- \frac{\partial}{\partial \hat{x}_1}\left(\hat{v}_{55}\frac{\partial \hat{\Phi}_3}{\partial \hat{x}_1} + \hat{v}'_{55}\frac{\partial \hat{\Phi}_3}{\partial \hat{x}_1}\right) - \frac{\partial}{\partial x_2}\left(v_{44}\frac{\partial \hat{\Phi}_3}{\partial x_2} + v'_{44}\frac{\partial \Phi_2}{\partial \hat{x}_3}\right) - \frac{\partial}{\partial \hat{x}_3}\left(\hat{v}_{11}\frac{\partial \hat{\Phi}_3}{\partial \hat{x}_3} + v_{12}\frac{\partial \Phi_2}{\partial x_2} + \hat{v}_{13}\frac{\partial \hat{\Phi}_1}{\partial \hat{x}_1}\right) = 0$$

Equations of state (modified Hooke's law) could be obtained from Eqs.(C.8), (C.10) and (C.12) in the following form:

$$\hat{u}_{11} = \hat{s}_{11}\hat{\sigma}_{11} + s_{12}\sigma_{22} + \hat{s}_{13}\hat{\sigma}_{33} + \hat{Q}_{11}\hat{P}_1^2 + Q_{12}P_2^2 + \hat{Q}_{13}\hat{P}_3^2 + \hat{R}_{11}\hat{\Phi}_1^2 + R_{12}\Phi_2^2 + \hat{R}_{13}\hat{\Phi}_3^2 + +\hat{F}_{11}\frac{\partial \hat{P}_1}{\partial \hat{x}_1} + F_{12}\frac{\partial P_2}{\partial x_2} + \hat{F}_{13}\frac{\partial \hat{P}_3}{\partial \hat{x}_3}$$

$$u_{22} = s_{12}\hat{\sigma}_{11} + s_{11}\sigma_{22} + s_{12}\hat{\sigma}_{33} + Q_{12}\hat{P}_1^2 + Q_{11}P_2^2 + Q_{12}\hat{P}_3^2 + R_{12}\hat{\Phi}_1^2 + R_{11}\Phi_2^2 + R_{12}\hat{\Phi}_3^2 + F_{11}\frac{\partial P_2}{\partial x_2} + F_{12}\left(\frac{\partial \hat{P}_1}{\partial \hat{x}_1} + \frac{\partial \hat{P}_3}{\partial \hat{x}_3}\right)$$

$$\hat{u}_{33} = \hat{s}_{13}\hat{\sigma}_{11} + s_{12}\sigma_{22} + \hat{s}_{11}\hat{\sigma}_{33} + \hat{Q}_{13}\hat{P}_1^2 + Q_{12}P_2^2 + \hat{Q}_{11}\hat{P}_3^2 + \hat{R}_{13}\hat{\Phi}_1^2 + R_{12}\Phi_2^2 + \hat{R}_{11}\hat{\Phi}_3^2 + \hat{F}_{11}\frac{\partial \hat{P}_3}{\partial \hat{x}_3} + F_{12}\frac{\partial P_2}{\partial x_2} + \hat{F}_{13}\frac{\partial \hat{P}_1}{\partial \hat{x}_1}$$

$$\hat{u}_{12} = \frac{1}{2}\left(s_{44}\hat{\sigma}_{12} + Q_{44}\hat{P}_1 P_2 + R_{44}\hat{\Phi}_1\Phi_2 + F_{44}\left(\frac{\partial \hat{P}_1}{\partial x_2} + \frac{\partial P_2}{\partial \hat{x}_1}\right)\right)$$

$$\hat{u}_{13} = \frac{1}{2}\left(\hat{s}_{55}\hat{\sigma}_{12} + \hat{Q}_{55}\hat{P}_1 \hat{P}_3 + \hat{R}_{55}\hat{\Phi}_1\hat{\Phi}_3 + \hat{F}_{55}\left(\frac{\partial \hat{P}_1}{\partial \hat{x}_3} + \frac{\partial \hat{P}_3}{\partial \hat{x}_1}\right)\right)$$

$$\hat{u}_{23} = \frac{1}{2}\left(s_{44}\hat{\sigma}_{23} + Q_{44}\hat{P}_3 P_2 + R_{44}\hat{\Phi}_3\Phi_2 + F_{44}\left(\frac{\partial P_2}{\partial \hat{x}_3} + \frac{\partial \hat{P}_3}{\partial x_2}\right)\right)$$

(C.16)

## APPENDIX D.

## Misfit strain impact on the anisotropic correlation length in thin BFO films

**Table II.** AFD and FE order parameter correlation lengths $L_C^\Phi$ and $L_C^P$ in BFO

| Order parameter | Type of the uncharged domain wall | | |
|---|---|---|---|
| | 180° | 109° | 71° |
| $\Phi_1$ | $L_C^\Phi = \sqrt{\dfrac{v_{11} + v_{12} + 2v_{44}}{-4\tilde{b}_1}}$ | Non applicable, since $\Phi_1 \approx const$ | Non applicable, since $\Phi_1 \approx const$ |
| $\Phi_2$ | $L_C^\Phi = \sqrt{\dfrac{v_{44}}{-2\tilde{b}_2}}$ | $L_C^\Phi = \sqrt{\dfrac{v_{44}}{-2\tilde{b}_2}}$ | |
| $\Phi_3$ | $L_C^\Phi = \sqrt{\dfrac{v_{11} - v_{12}}{-4\tilde{b}_1}}$ | $L_C^\Phi = \sqrt{\dfrac{v_{44}}{-2\tilde{b}_1}}$ | $L_C^\Phi = \sqrt{\dfrac{v_{44}}{-2\tilde{b}_1}}$ |
| $P_1$ | $L_C^P = \sqrt{\dfrac{g_{11} + g_{12} + 2g_{44}}{-4\tilde{a}_1}}$ | Non applicable, since $P_1 \approx const$ | Non applicable, since $P_1 \approx const$ |
| $P_2$ | $L_C^P = \sqrt{\dfrac{g_{44}}{-2\tilde{a}_2}}$ | $L_C^P = \sqrt{\dfrac{g_{44}}{-2\tilde{a}_2}}$ | Non applicable, since $P_2 \approx const$ |
| $P_3$ | $L_C^P = \sqrt{\dfrac{g_{11} - g_{12}}{-4\tilde{a}_1}}$ | $L_C^P = \sqrt{\dfrac{g_{44}}{-2\tilde{a}_1}}$ | $L_C^P = \sqrt{\dfrac{g_{44}}{-2\tilde{a}_1}}$ |

Here used the following designations for coefficients dependence on the misfit strain:

$$\tilde{a}_1 = a_1 - \frac{Q_{11} + Q_{12}}{s_{11} + s_{12}}u_m, \quad \tilde{a}_2 = a_1 - \frac{2Q_{12}}{s_{11} + s_{12}}u_m \qquad (D.1a)$$

$$\tilde{b}_1 = b_1 - \frac{R_{11}+R_{12}}{s_{11}+s_{12}} u_m, \quad \tilde{b}_2 = b_1 - \frac{2R_{12}}{s_{11}+s_{12}} u_m \tag{D.1b}$$

The figure S1 shows what happens if one increases the gradient coefficients in such a way to approach the gradient instability threshold, when the width of the domain walls increases significantly.

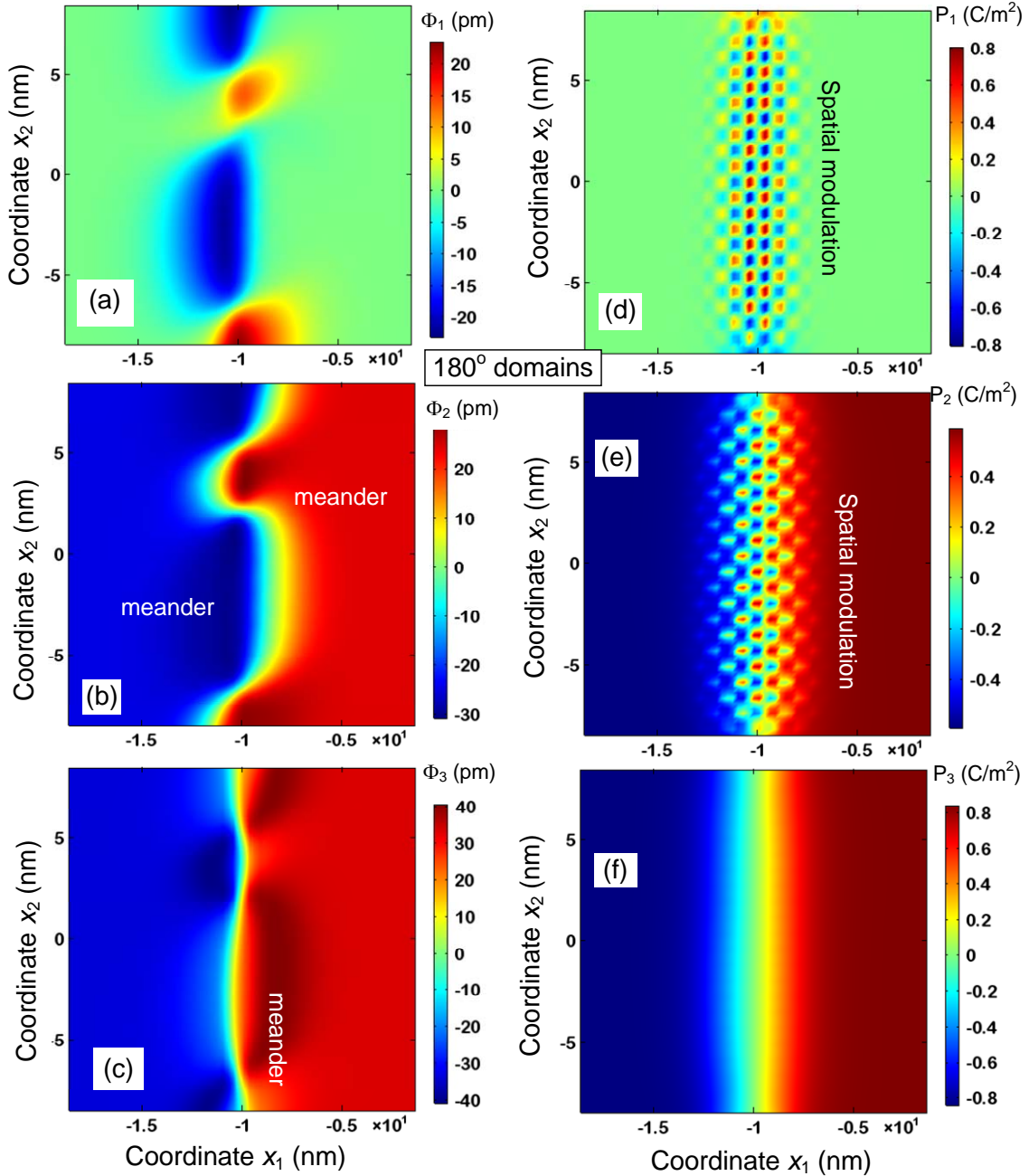

**FIGURE S1| {101} oriented 180° DWs in 18 nm BFO film on PSO substrate, after a very long relaxation time.** Distribution of the AFD order parameter $\Phi_i$ **(a)-(c)** and polarization $P_i$ **(d)-(f)** components in the $\{x_1 x_2\}$ cross-section of a thin BFO film with {101} oriented 180° DWs. Film thickness is 18 nm, room temperature. Tilted coordinate system is used for $x_1$. Polarization gradient coefficients are $g_{11}=27$, $g_{12}=-27$, $g_{44}=27$ (in $10^{-10} C^{-2} m^3 J$) and tilt gradient coefficients $v_{11}=1$, $v_{12}=0.97$ and $v_{44}=2$ (in $10^{11}$ J/m$^3$).

# APPENDIX E.

## Estimates of local static conductivity at 180° DWs in BFO films

The concentration of free electrons $n$ in the conductive band in a BFO film of thickness more than 10 nm can be estimated in the continuous levels approximation [iv]:

$$n(\varphi, u_{ij}) = \int_0^\infty d\varepsilon \cdot g_n(\varepsilon) f(\varepsilon - E_F + E_C(u_{ij}) - e\varphi), \quad (E.1)$$

Where $g_n(\varepsilon) \approx \sqrt{2m_n^3 \varepsilon}/(2\pi^2 \hbar^3)$ is the density of states in the effective mass approximation, $f(x) = (1 + \exp(x/k_B T))^{-1}$ is the Fermi-Dirac distribution function, $k_B = 1.3807 \times 10^{-23}$ J/K, $T$ is the absolute temperature and $e = 1.6 \times 10^{-19}$ C is the electron charge. $E_F$ is the Fermi energy level. The strain-dependent bottom of conductive band is

$$E_C(u_{ij}) = E_{C0} + \Xi_{ij}^C \delta u_{ij}. \quad (E.2a)$$

Where $\Xi_{ij}^C$ is the deformation potential tensor, $\delta u_{ij} = u_{ij} - u_{ij}^S$; value $E_{C0}$ already includes the spontaneous strain $u_{ij}^S$ existing far from the domain wall. Notably, that

$$u_{ij} = -\frac{\delta G}{\delta \sigma_{kl}} \cong s_{ijkl}\sigma_{kl} + Q_{ijkl}P_k P_l + R_{ijkl}\Phi_k \Phi_l + F_{ijkl}\frac{\partial P_k}{\partial x_l} \quad (E.2b)$$

in accordance with Eq. (C.16). So that the strain contains the contribution from electrostriction, rotostriction and flexoelectric effect. The estimations for the band gap derivative $\partial E_g / \partial u_{ij}$ ~20 eV and $|\Xi_{ij}^{C,V}|$ ~ 20 eV are available for e.g. BiFeO$_3$ [v]. The concentration of holes in the valence band can be considered in a similar way.

The local static conductivity, directly related with CAFM contrast, can be estimated as

$$\sigma = e\eta_e n. \quad (E.3)$$

The mobility of holes is regarded negligible in comparison with electron mobility, $\eta_e \gg \eta_h$, and the conductivity in Eq.(E.3) is purely of n-type. Thus, in accordance with the above theoretical estimates, the experimentally observed conductivity enhancement at the domain wall can be caused by the electric potential and elastic strain variation inside the wall. The electro-elastic potential relief (wells or humps) leads to higher concentration of electrons (or holes) in the DW due to the local band bending proportional to $E_C(u_{ij}) - e\varphi$, per Eq.(E.2). It appeared that the contribution of the strain variations to the conductivity modulation is dominant, because the potential is zero exactly at the 180° and 71° DWs. Since the strain variation (E.2b) is proportional to $Q_{ijkl}P_k P_l + R_{ijkl}\Phi_k \Phi_l + F_{ijkl}\frac{\partial P_k}{\partial x_l}$, it may also include the atomic bond changes at the wall via the striction and flexoelectric mechanisms.

Figure S2 shows the map of normalized static conductivity (that is directly proportional to CAFM contrast) across {101} oriented 180° DWs in 18 nm BFO film on PSO substrate. From the figure the DW is one order of magnitude more conductive than the domains.

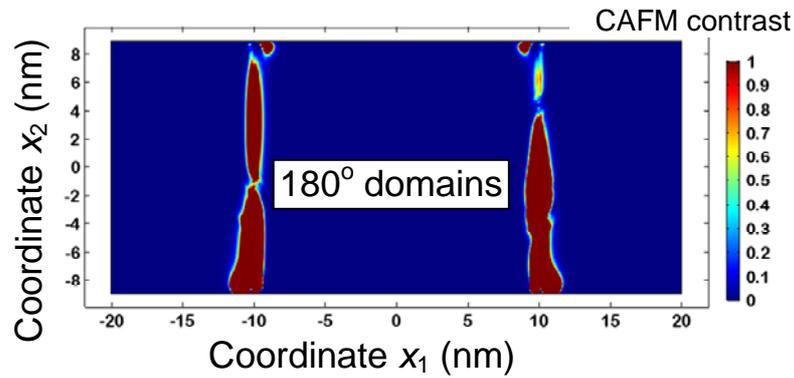

**FIGURE S2| Normalized static conductivity across {101} oriented 180° DWs in 18 nm BFO film on PSO substrate.** Parameters are listed in Table I, the trace $\Xi_{ii}^{C}$ = 20 eV.